\documentclass[a4paper]{article}
\usepackage{a4wide}
\usepackage[hyphens]{url}
\usepackage{siunitx,booktabs,subcaption}
\usepackage{tabularx}
\usepackage{grffile}
\usepackage{authblk}
\usepackage{amsmath}
\usepackage{enumerate}
\usepackage{amssymb}
\usepackage{lmodern}
\usepackage{xfrac}
\usepackage{multirow}% http://ctan.org/pkg/multirow
\newcommand{\ra}[1]{\renewcommand{\arraystretch}{#1}}
\DeclareSIPrePower\quartic{4}
\DeclareSIPostPower\tothefourth{4}

\usepackage[numbers]{natbib}
\graphicspath{{./figures}}

\usepackage{csquotes}
\usepackage[UKenglish,USenglish]{babel}
\usepackage{hyperref}

\title%[CoA impact upon LV load]%
{Towards a Computational Framework for Modeling the Impact of Aortic Coarctations upon Left Ventricular Load}
\author[1]{Elias Karabelas}
\author[1]{Matthias A. F. Gsell}
\author[1,2]{Christoph M. Augustin}
\author[1]{Laura Marx}
\author[1]{Aurel Neic}
\author[1]{Anton J. Prassl}
\author[3,4]{Leonid Goubergrits}
\author[3,4]{Titus Kuehne}
\author[$1,\ast$]{Gernot Plank}
\affil[1]{Institute of Biophysics, Medical University of Graz, Graz, Austria}
\affil[2]{Shadden Research Group, Department of Mechanical Engineering, University of California, Berkeley, CA, USA}
\affil[3]{Department of Congenital Heart Disease/Pediatric Cardiology, German Heart Institute Berlin, Berlin, Germany}
\affil[4]{Institute for Imaging Science and Computational Modelling in Cardiovascular Medicine, Charit\'{e}-Universit\"atsmedizin Berlin, Berlin, Germany}
\affil[$\ast$]{Correspondance: Gernot Plank, Medical University of Graz,
  Institute of Biophysics, Neue Stiftingtalstrasse 6, 8010 Graz, Austria. gernot.plank@medunigraz.at}
\date{}

\renewcommand{\vec}[1]{\boldsymbol{#1}}

\renewcommand{\vec}[1]{\mathbf{#1}}

\newcommand{\abs}[1]{\left\lvert{#1}\right\rvert}
\newcommand{\tens}[1]{\boldsymbol{\mathsf{#1}}}
\newcommand{\dt}{\,\mathrm{d}t}
\newcommand{\dd}{\,\mathrm{d}}
\newcommand{\dx}{\, \mathrm{d}\vec{x}}
\newcommand{\dsx}{\, \mathrm{d}s_{\vec{x}}}

\newcommand{\tr}{\operatorname{tr}}
\newcommand{\partd}[2]{\frac{\partial {#1}}{\partial {#2}}}

\newcommand{\tensor}[1]{\boldsymbol{\mathbf{#1}}}

\newcommand{\isotensor}[1]{\overline{\boldsymbol{\mathbf{#1}}}}

\newcommand{\einschraenkung}[1]{\big|_{#1}}

\begin{document}
% Full document: 12000 words

\maketitle

\begin{abstract}
% Abstract: 350 words
Computational fluid dynamics (CFD) models of blood flow in the left ventricle (LV)
and aorta are important tools for analyzing the mechanistic links between
myocardial deformation and flow patterns.
Typically, the use of image-based kinematic CFD models prevails in applications
such as predicting the acute response to interventions which alter LV afterload conditions.
However, such models are limited in their ability to analyze any impacts upon LV load
or key biomarkers known to be implicated in driving remodeling processes
as LV function is not accounted for in a mechanistic sense.

This study addresses these limitations by reporting on progress made towards a
novel electro-mechano-fluidic (EMF) model
that represents the entire physics of LV electromechanics (EM) based on first principles.
A biophysically detailed finite element (FE) model of LV EM
was coupled with a FE-based CFD solver for moving domains
using an arbitrary Eulerian-Lagrangian (ALE) formulation.
Two clinical cases of patients suffering from aortic coarctations (CoA)
  were built and parameterized based on clinical data under pre-treatment conditions.
For one patient case simulations under post-treatment conditions
after geometric repair of CoA by a virtual stenting procedure
  were compared against pre-treatment results.
Numerical stability of the approach was demonstrated by analyzing mesh quality
and solver performance under the significantly large deformations of the LV  blood pool.
Further, computational tractability and compatibility with clinical time scales
were investigated by performing strong scaling benchmarks up to 1536 compute cores.
The overall cost of the entire workflow for building, fitting and executing EMF simulations
was comparable to those reported for image-based kinematic models, suggesting
that EMF models show potential of evolving into a viable clinical research tool.

{\textbf{Keywords:} cardiac mechanics, computational fluid dynamics,
 	finite element model, arbitrary Lagrangian-Eulerian formulation, patient-specific modeling,
 	translational cardiac modeling, total heart function.}
\end{abstract}
%==============================================================================
\section{Introduction}
CFD models of blood flow in the LV and aorta are important tools for analyzing
the mechanistic links between myocardial deformation and flow patterns.
Typically, such models are either driven by prescribed flow profiles measured
in the LV outflow tract or the aortic root
\cite{andersson2017multidirectional,goubergrits2013impact,ralovich2015noninvasive},
or by image-based kinematic models
\cite{doenst2009fluid, Chnafa2014, schenkel2009mri, seo2013multiphysics,%
mihalef2011patient, su2016cardiac}
built from segmentation of 4D medical imaging datasets.
While such models have proven to be valuable for analyzing the hemodynamic
status quo of a patient or for predicting changes in hemodynamics in the aorta
secondary to intervention such as aortic valve repair \cite{kelm2017model} or
stenting of a coarctation \cite{goubergrits2015mri},
they are inherently limited in their ability to assess cardiac function
as the biophysics driving myocardial activation and deformation is not taken
into consideration in the model formulation.
EMF models that capture the entire physics of a heartbeat based on first
principles show promise to overcome this limitation \cite{crozier2016imagebased}
by rendering feasible the assessment of all essential myocardial parameters,
which are known to be key factors driving ventricular remodeling and disease progression.
Thus EMF models may offer, in principal, the potential of predicting longer term outcomes
beyond changes in the acute response to therapies.

However, due to a number of factors such as the inherent complexity of multiphysics models,
the large-scale motion and complex deformation of the myocardial walls
as well as the significant computational burden,
these models pose substantial methodological challenges.
%%%
For LV EMF models and similar applications, methods to overcome the problem of
large-scale deformations can be roughly
classified into two categories: ALE formulations using a moving fluid mesh
\cite{devecchi2016novel,nordsletten2011fluid,Tang2010,tang2008patient,vazquez2015alya} and
immersed boundary (IB) methods \cite{choi2015anew,seo2013effect,vigmond2008effect}.
While ALE formulations often rely on severe simplifications or automatic remeshing strategies
\cite{long2013fluid}, IB methods are more versatile as the
moving wall of the ventricle is not explicitly tracked.
However, IBs and all related non-boundary-fitting methods have a reduced accuracy for the
solution near the fluid-solid structure interface due to interpolation errors,
pose severe challenges on the implementation,
and additional degrees of freedom have to be introduced on interface cut elements,
which all contributes to significantly higher computational costs
\cite{vanloon2007comparison}.

In this study, we report on the progress made towards a novel EMF model of the human LV
that is entirely based on first principles and that copes with significantly
large defomations, i.e., ejection fractions (EFs) beyond 60 \%, without requiring remeshing or IB principles.
Validated \emph{in silico} models taken from a recent clinical modeling study
where a cohort of \emph{in silico} EM LV and aorta models
of patients suffering from aortic valve disease (AVD)
and/or CoA \cite{augustin16:_patient} were built,
served as kinematic driver to a computational model of hemodynamics in the LV cavity and aorta.
A hybrid two stage modeling approach was adopted with regard to hemodynamics.
First, the afterload imposed by the circulatory system onto the LV was represented
by a lumped model of afterload and coupled to an EM model of LV and aorta
to compute LV kinematics.
Subsequently a full-blown CFD model with moving domain boundaries
based on an ALE formulation
was \emph{unidirectionally} or \emph{weakly} coupled to the EM model
using the kinematics of its endocardial surface as input.
We show validation results for two selected clinical CoA cases under pre-treatment conditions
and compare pre-treatment and post-treatment simulation results for one patient case
in which the CoA was geometrically repaired by a virtual stenting procedure.
Further, we demonstrate numerical feasibility of the implemented approach
by analyzing changes in mesh quality and its impact upon solver performance
under the significantly large deformations of the LV blood pool mesh
and also provide strong scaling benchmarking results for a range of 96 to 1536 compute cores.
The overall cost of the entire workflow for building, fitting and execution of EMF simulations
is $\approx$ 48 hours which is comparable to plain image-based kinematic driver
models \cite{mittal16:_computational}.

%==============================================================================
\section{Methods}
The methodology to develop a coupled model of cardiac and cardiovascular
hemodynamics based on an ALE formulation is structured as follows.
\begin{enumerate}[i) ]
  \item We begin in Section~\ref{sec:ClinicalDataAcquisition} by describing
    MRI data acquisition and anatomical FE model generation of
    the LV and aorta for two patients suffering from CoA.
  \item Then, a brief summary of all model components is given comprising
    an electrophysiology (EP) model to drive electrical activation and
    repolarization (Section \ref{sec:eikonal});
    an EM model describing passive biomechanics as well as the generation of
    active stresses (Section \ref{sec:TissueMechanicsModel});
    afterload models to provide appropriate boundary conditions on the LV
    endocardium during the ejection phase
    (Section \ref{sec:afterload_models});
    and a CFD model with moving domain boundaries
    representing blood flow in the LV and aorta during ejection.
    The EM and CFD model are weakly coupled in a forward fluid structure
    interaction (FSI) framework, where the EM model is used as a kinematic
    driver to move the fluid domain (Section~\ref{sec:FluidFlowModel}).
  \item The solution procedure and software implementation details are
    outlined in Section~\ref{sec:NumericalSolution}.
  \item Finally, procedures implemented for the patient-specific
    parameterization of the major model components is described in
    Section \ref{sec:meth:parameterization}.
\end{enumerate}
%
%
%------------------------------------------------------------------------------
\subsection{Clinical data acquisition and model generation}
\label{sec:ClinicalDataAcquisition}
Hemodynamic data of two patients with clinical indication for catheterization
due to CoA -- all preceding a cardiac magnetic resonance study -- were acquired
before and after CoA treatment by stent implant, see \autoref{tab:patients}.
CoA treatment indicators included an echocardiographic measured, peak systolic
pressure gradient across the stenotic region of $>\SI{20}{\mmHg}$ and/or
arterial hypertension.
The study was approved by the institutional research ethics committee
following the ethical guidelines of the 1975 Declaration of Helsinki.
Written informed consent was obtained from the participants' guardians.
Acquired data are summarized in \autoref{tab:patients}.
%
%------------------------------------------------------------------------------
\subsubsection{MRI acquisition and post processing}
\label{sec:MRIAcquisition}
MR imaging was done with a whole body 1.5 Tesla MR scanner Achieva R 3.2.2.0
using a five-element cardiac phased-array coil (Philips Medical System, Best,
Netherlands).
Three MRI sequences were used further in our study:
i) flow-sensitive four-dimensional (4D) velocity-encoded magnetic resonance
  imaging (4D VEC-MRI),
ii) three-dimensional (3D) anatomical imaging of the whole heart (3DWH)
  during diastasis, and
iii) 4D gapless short axis Cine MRI.

4D VEC-MRI of the thorax was performed using an anisotropic 4D segmented
$k$-space phase contrast gradient echo sequence.
Retrospective electrocardiographic gating without navigator gating of
respiratory motion in order to minimize acquisition time was used.
Sequence parameters were:
acquired voxel $2.5\times2.5\times\SI{2.5}{\milli\metre}$;
reconstructed voxel $1.7\times1.7\times\SI{2.5}{\milli\metre}$;
repetition time $\SI{3.5}{\milli\second}$;
echo time $\SI{2.2}{\milli\second}$;
flip angle $\ang{5}$; 25 reconstructed cardiac phases;
number of signal averages 1;
High velocity encoding (\SIrange{3}{6}{\meter\per\second})
in all three directions was used in order to avoid phase wraps in the presence
of coarctation and associated secondary flow. Flow measurements were completed
with automatic correction of concomitant phase errors.
Postprocessing for analysis of flow rates across the aortic valve was carried
out with GTFlow 1.6.8 software%
\footnote{\url{http://www.gyrotools.com/products/gt-flow.html}}
(Gyrotools, Zurich, Switzerland).

The 3DWH exemplary sequence parameters were:
acquired voxel $0.66\times0.66\times\SI{3.2}{\milli\metre}$;
reconstructed voxel $0.66\times0.66\times\SI{1.6}{\milli\metre}$;
repetition time $\SI{4.0}{\milli\second}$;
echo time $\SI{2.0}{\milli\second}$;
flip angle $\ang{90}$; and
 number of signal averages 3.

Short axes Cine imaging data were acquired with sequence parameters:
16 slices, with an acquisition resolution of $0.86 \times 0.86 \times 6.0$ mm,
repetition time 4.24~ms,
echo time 2.12~ms,
flip angle 60$^\circ$ and 25 automatically reconstructed cardiac phases
which were used to determine LV volume traces.
The non-compact myocardium as well as papillary muscles
were counted towards blood pool volume.

MRI based pressure mapping allowing to assess non-invasively the relative
pressures in a vessel by solving Pressure Poisson equation (PPE) was done with
MevisFlow\footnote{\url{https://www.mevis.fraunhofer.de/en/solutionpages/%
mevisflow-non-invasive-interactive-exploration-of-in-vivo-hemodynamics.html}}.
Briefly, the PPE can be derived from the Navier--Stokes equations
by taking the divergence of the momentum equation \eqref{eq:navier_stokes_pointwise:pde:1},
see \cite{gresho1987pressure, KRITTIAN20121029} for more details.
The processing and analysis pipeline of the pressure mapping consists of the
following four steps.
\begin{enumerate}[i) ]
  \item Semi-automatic segmentation (labeling) of the aortic domain from 3DWH
    data generating 3D mask of the aorta.
  \item Background phase correction and phase-unwrapping of the 4D VEC-MRI data
     and generation of a sequence of volumetric velocity vector fields.
  \item Coarse semi-automatic segmentation of the aorta based on magnitude and
     phase contrast of the 4D VEC-MRI data and registration with 3DWH based
     mask of the aorta.
   \item Solving the PPE at each time step having 4D VEC-MRI data as input. Furthermore, a \SI{5}{\percent} mask size reduction is applied in order to
     avoid numerical inconsistencies close to the vessel wall as suggested
     earlier \cite{meier2010PCMRI}.
\end{enumerate}
Relative pressure maps are represented with zero pressure located at the center
of the CoA (narrowest location). 3D mask based on 3DWH data was used due to
its better spatial resolution compared to 4D VEC-MRI data.
Correction of velocity data (step ii) was done in order to minimize noise and
aliasing artifacts originating from multiple sources.
%
%------------------------------------------------------------------------------
\subsubsection{Invasive catheter recordings}
During catheterization, pressure was recorded over the cardiac cycle
in the ascending aorta and the LV before treatment and repeated in the
ascending aorta after an interventional treatment procedure was performed.
Pressures were recorded simultaneously at three predefined locations (LV,
ascending aorta, and descending aorta) and the femoral artery during catheterization.
Patients were sedated by intravenous administration of a bolus of midazolam
(\SIrange[]{0.1}{0.2}{\mg\per\kg}, max. \SI{5}{\mg}),
followed by a bolus of propofol (\SIrange{1}{2}{\mg\per\kg}, as needed)
and continuous infusion of propofol (approximately \SI{4}{\mg\per\kg/\hour},
 as needed).
Pressure measurements were taken with senior cardiologists present.
Pigtail catheters (Cordis, Warren, NJ, USA) of 5-6F were connected to
pressure transducers (Becton-Dickinson, Franklin Lakes, NJ, USA).
Routinely, patients received balloon angioplasty with or without additional
placement of a stent in order to treat a given stenosis by removing the
narrowing of the vessel and thus the pressure gradient.
To reduce duration of catheterization, pressures were measured post-treatment
only in the ascending aorta. The Schwarzer hemodynamic analysis system
(Schwarzer, Heilsbronn, Germany) was used
to amplify, acquire, and analyze pressure signals.

%
%------------------------------------------------------------------------------
\subsubsection{Anatomical FE Model Generation}\label{sec:anatomic_model_generation}
\begin{figure}[htb]
  \centering
  \includegraphics[width=\textwidth,keepaspectratio]{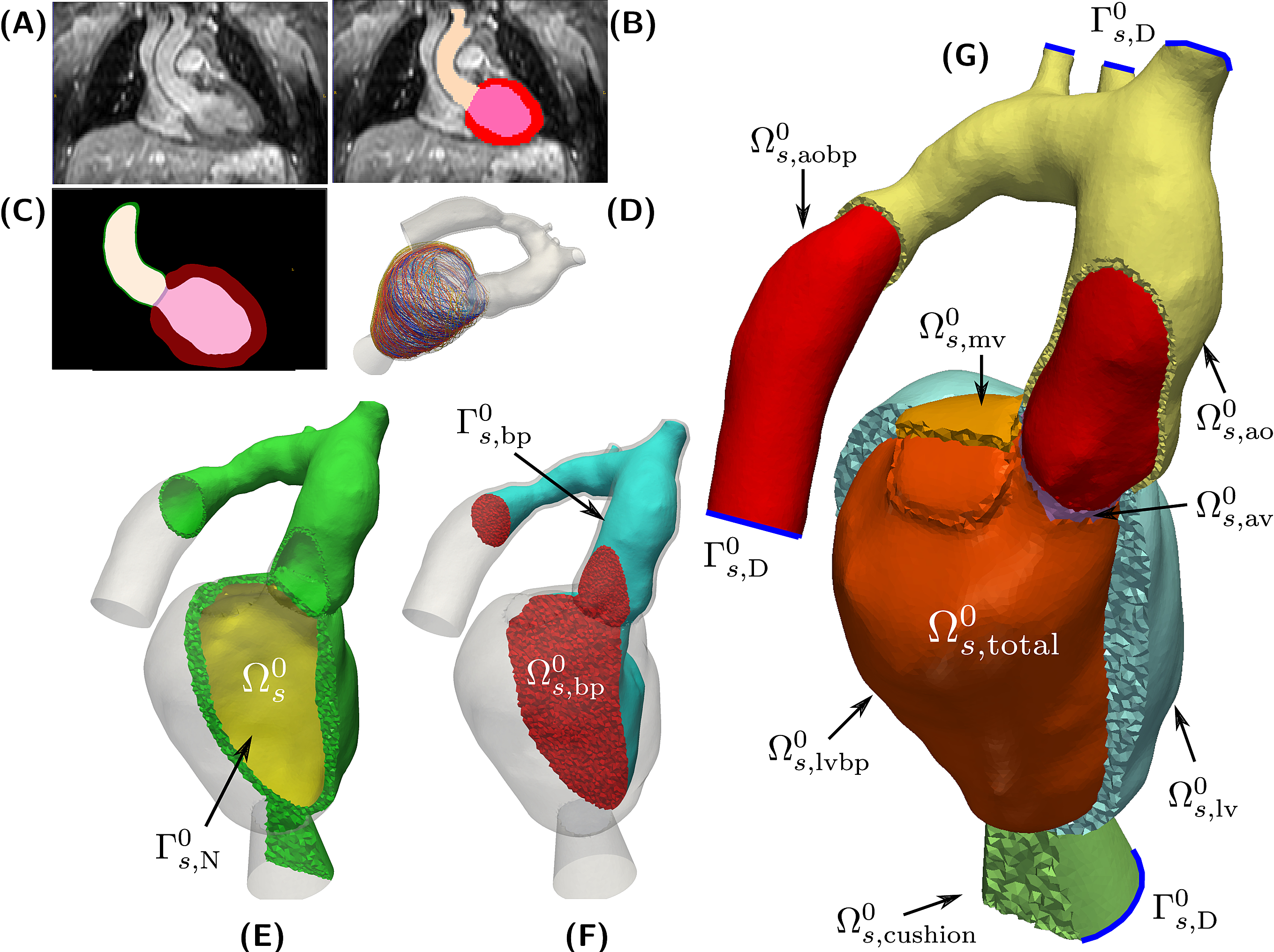}
  \caption{Mechanics model generation: Starting from a patient specific MRI scan
    \textsf{(A)} a segmentation was performed \textsf{(B)} which was then
    upsampled and smoothed \textsf{(C)}. Myocardial fibers were generated
    in the tissue according to \cite{bayer2012novel} \textsf{(D)}.
    A labeled FE geometry $\Omega_{\mathrm{s,total}}^0$ including
    the blood pool was generated \textsf{(G)}.
    The geometry has been sliced to reveal the blood pool and valves and has
    been color coded according to the labels defined in \autoref{tab:taglabels}.
    Boundaries $\Gamma_{\mathrm{s,D}}^0$ used for prescribing homogeneous
    Dirichlet boundary conditions are sketched as blue curves.
    From this mesh the
    EM submesh $\Omega_\mathrm{s}^0$ \textsf{(E)}
    and the unsmoothed blood pool \textsf{(F)} were extracted.
    Boundary $\Gamma_{\mathrm{s,N}}^0$ was used to prescribe pressure
    boundary conditions inside the LV and $\Gamma_{\mathrm{s,bp}}^0$
    is the surface of the blood pool.
    %The lower part of this figure compares the cases
    %from \autoref{tab:patients}. The coarctations are marked with green circles.
    % For the cases \emph{28-Pre} and \emph{28-Post} a visual comparison is
    %depicted showing the effect of stenting.}\label{fig:mechanics_model_generation}
  }
  \label{fig:mechanics_model_generation}
\end{figure}

\begin{figure}[htb]
  \centering
  \includegraphics[width=\textwidth,keepaspectratio]{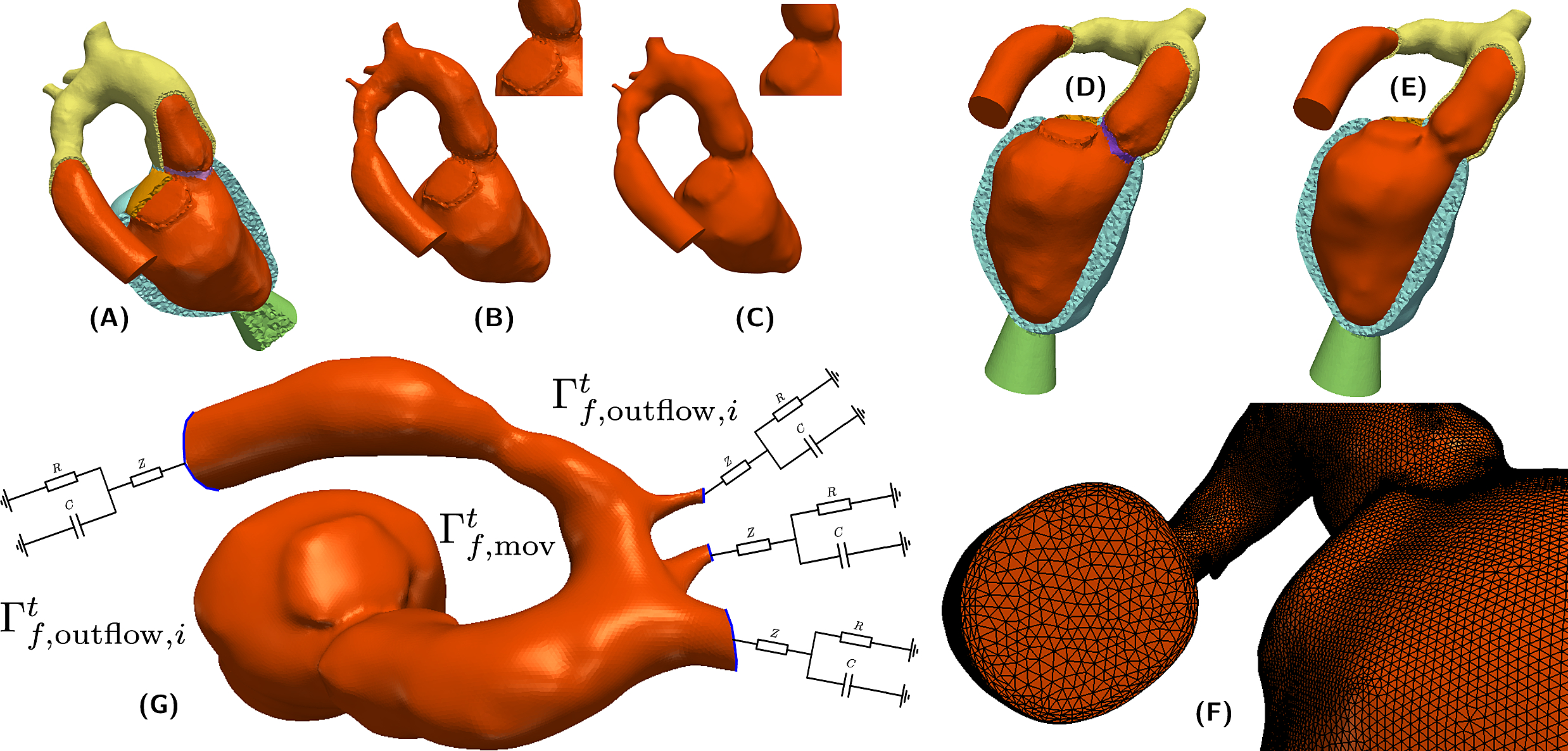}
  \caption{Processing workflow used for generating blood pool FE models:
      \textsf{(A)} and \textsf{(B)} Elements labeled as blood pool or valve
      were extracted from the mesh used for EM modeling.
      \textsf{(C)} Surfaces of extracted meshes were smoothed to avoid numerical
      instabilities due to reentrant corners resulting from a jagged surface.
      A closeup view of the smoothing effect is displayed in the upper right.
      The smoothed surface is then used as input for the fluid mesh generation.
      \textsf{(D)} and \textsf{(E)} Comparison of the smoothed and unsmoothed
      blood pool mesh immersed in the original EM mesh.
      \textsf{(F)} Closeup view of the generated boundary layer mesh.
      \textsf{(G)} Boundary conditions used for CFD.
      Moving wall boundary $\Gamma_{\mathrm{f,mov}}^t$ colored in orange,
      outlet boundaries $\Gamma_{\mathrm{f,outflow},i}^t$ colored in blue
      with attached illustration of the 3-element Windkessel models.
  }
  \label{fig:fluid_model_generation}
\end{figure}

Multi-label segmentation of the LV myocardium, LV blood pool, left atrium (LA)
and aortic cavities was done at the DHZB using 3DWH data and the
ZIB Amira software\footnote{\url{https://amira.zib.de}} \cite{stalling2005amira}.
The segmentations were smoothed and upsampled to a \SI{0.1}{\milli\metre}
isotropic resolution using a variational smoothing method \cite{crozier2016imagebased}.
The resulting high resolution multi-label segmentation was meshed using
CGAL\footnote{\url{http://www.cgal.org}} \cite{cgal:eb-17b},
giving a global mesh $\Omega_{\mathrm{s},\mathrm{total}}^0$ consisting of
tetrahedral elements.
Here, $(\bullet)^0$ denotes the mechanical reference configuration at
end-diastolic pressure. The mesh was subdivided into various subdomains
corresponding to predefined labels which are summarized in \autoref{tab:taglabels}.
We write
\begin{equation}
    \Omega_{\mathrm{s},\mathrm{total}}^0 =
    \bigcup_{i \in I} \Omega_{\mathrm{s},i}^0,
\end{equation}
with the index set
\begin{equation}
  I := \{\mathrm{lv},\mathrm{ao},\mathrm{cushion},\mathrm{av},\mathrm{mv},
         \mathrm{lvbp},\mathrm{aobp}\},
\end{equation}
see \autoref{fig:mechanics_model_generation} \textsf{(E--G)} for illustration.
With this, we define the following submeshes
\begin{align}
  \Omega_\mathrm{s}^0 &:= \Omega_{\mathrm{s,total}}^0 \backslash
    \left(\Omega_{\mathrm{s,lbvp}}^0 \cup
    \Omega_{\mathrm{s,aobp}}^0\right),\\
  \Omega_\mathrm{s,bp}^0 = \widetilde{\Omega}_{f}^0&:=
    \Omega_{\mathrm{s,av}}^0 \cup
    \Omega_{\mathrm{s,lvbp}}^0 \cup \Omega^0_{\mathrm{s,aobp}},
%\Omega_{s,\mathrm{lv}}^0 &:= \Omega_{s,\mathrm{lv}}^0.
\end{align}
where $\Omega_\mathrm{s}^0$ is the solid domain and $\Omega_\mathrm{s,bp}^0$
is the unsmoothed blood pool domain used for extracting a smoothed CFD mesh,
see \autoref{fig:mechanics_model_generation} \textsf{(E)} and
\textsf{(F)}.
For later use, we define the following surfaces
\begin{align}
  \Gamma_{\mathrm{s,N}}^0 &:= \partial\left(\left(\Omega_{s,\mathrm{lv}}^0 \cup
    \Omega_{\mathrm{s,av}}^0 \cup
    \Omega_{\mathrm{s,mv}}^0\right) \cap
    \Omega_{\mathrm{s,lvbp}}^0\right),\\
%  \widehat{\Gamma}_{\mathrm{D},i} &:=
%    \partial \left(\widehat{\Omega}_\text{EM} \cap \varepsilon_i\right),
%    \qquad \text{for } i=0,\ldots,2,\\
%  \widehat{\Gamma}_{\mathrm{D}} &:=
%    \bigcup_{i=0}^{2} \widehat{\Gamma}_{\mathrm{D},i},\\
 \Gamma_{\mathrm{s,H}}^0 &:= \partial \Omega_\mathrm{s}^0 \backslash
    \left(\Gamma_{\mathrm{s,N}}^0 \cup \Gamma_{\mathrm{s,D}}^0 \right),\\
 \Gamma_{\mathrm{s,bp}}^0 &:=
    \partial \Omega_{\mathrm{s,bp}}^0 \backslash \Gamma_{\mathrm{s,D}}^0,
\end{align}
where $\Gamma_{\mathrm{s,D}}^0$ denote the cutoff faces as indicated by blue
lines in \autoref{fig:mechanics_model_generation};
$\Gamma_{\mathrm{s,N}}^0$ are surfaces subject to pressure; and
$\Gamma_{\mathrm{s,H}}^0$ are surfaces with homogeneous Neumann boundary
conditions.
In order to avoid numerical difficulties with non-smooth, jagged boundaries,
the surface of the mechanical blood pool domain $\Gamma_{\mathrm{s,bp}}^0$
was extracted and smoothed using the
VMTK toolbox\footnote{\url{http://www.vmtk.org}} \cite{antiga2008image}.
The smoothed surface, $\Gamma_{\mathrm{f,wall}}^0$, was used to define the
boundary of the fluid domain reference configuration, $\Omega_\mathrm{f}^0$,
for volumetric FE meshing using
ANSYS ICEM CFD\footnote{\url{http://www.ansys.com/Services/training-center/
platform/introduction-to-ansys-icem-cfd-Hexa}}.
Refined boundary layers were included in this process to better resolve sharp
gradients in the vicinity of $\Gamma_{\mathrm{f,wall}}^0$ occurring during
simulation of hemodynamics.
The various processing stages for building EM and CFD models are illustrated
in Figures~\ref{fig:mechanics_model_generation} and
\ref{fig:fluid_model_generation}, respectively.
%
%------------------------------------------------------------------------------
\subsection{Electromechanical Model}
\subsubsection{Electrophysiology of the LV}
\label{sec:eikonal}
A recently developed reaction-eikonal (R-E) model \cite{neic17:_efficient} was
employed to generate electrical activation sequences
which serve as a trigger for active stress generation in cardiac tissue.
The hybrid R-E model combines a standard reaction-diffusion (R-D) model
based on the monodomain equation with an eikonal model.
Briefly, the eikonal equation is given as
\begin{equation} \label{eq:_eikonal}
  \left\{
    \begin{array}{rcll}
      \sqrt{\nabla_{\vec X} t_\mathrm{a}^\top \, \tensor{V} \,
      \nabla_{\vec X} t_\mathrm{a}} & =
      & 1 \qquad & \text{in }  \Omega_{\mathrm{s,lv}}^0, \\
      t_\mathrm{a} & = & t_0 & \text{on } \Gamma_{\mathrm{s},\ast}^0,
    \end{array}
  \right.
\end{equation}
where $(\nabla_{\vec X})$ is the gradient with respect to the end-diastolic
reference configuration $\Omega_{\mathrm{s,lv}}^\mathrm{0}$;
$t_\mathrm{a}$ is a positive function describing the wavefront arrival
time at location $\vec{X} \in \Omega_{\mathrm{s,lv}}^\mathrm{0}$;
and $t_0$ are initial activations at locations
$\Gamma_{\mathrm{s},\ast}^0 \subseteq \Gamma_{\mathrm{s,N}}^\mathrm{0}$.
%\todoCA{What is $\Gamma_{\mathrm{s},\ast}^0$?}
The symmetric positive definite $3 \times 3$ tensor $\tensor{V}(\vec{X})$
holds the squared velocities
$\left(v_\mathrm{f}(\vec{X}),v_\mathrm{s}(\vec{X}),v_\mathrm{n}(\vec{X})\right)$
associated to the tissue's eigenaxes,
referred to as fiber, $\vec{f}_0$, sheet, $\vec{s}_0$,
and sheet normal, $\vec{n}_0$, orientations. The arrival time function
$t_\mathrm{a}(\vec{X})$ was subsequently used in a modified
monodomain R-D model given as
\begin{equation} \label{equ:_monodomain_R_E}
  \beta C_\mathrm{m} \frac{\partial V_\mathrm{m}}{\partial t} =
    \nabla_{\vec X} \cdot \tensor{\sigma}_\mathrm{i} \nabla_{\vec X} V_\mathrm{m} +
    I_\mathrm{foot} - \beta I_\mathrm{ion},
\end{equation}
where an arrival time dependent foot current,
$I_{\mathrm{foot}}(t_\mathrm{a})$, was added
which is designed to mimic subthreshold electrotonic currents
to produce a physiological foot of the action potential.
The key advantage of the R-E model is its ability
to compute activation sequences at much coarser spatial resolutions
that are not afflicted by the spatial undersampling artifacts
leading to conduction slowing or even numerical conduction block as it is
observed in standard R-D models.
Ventricular EP was represented by the tenTusscher--Noble--Noble--Panfilov model
of the human ventricular myocyte \cite{tentusscher04:_TNNP}.
As indicated in Equations~(\ref{eq:_eikonal},~\ref{equ:_monodomain_R_E}),
activation sequences and electrical source distribution in the LV were computed
in its end-diastolic configuration $\Omega_{\mathrm{s,lv}}^\mathrm{0}$,
that is, any effects of deformation upon electrotonic currents remained unaccounted for.
%
%------------------------------------------------------------------------------
\subsubsection{Active and Passive Mechanics in the LV and Aorta}
\label{sec:TissueMechanicsModel}
The deformation of the heart is governed by imposed external loads
such as pressure in the cavities or from surrounding tissue
and active stresses intrinsically generated during contraction.
Tissue properties of the LV myocardium and the aorta are characterized as a
hyperelastic, nearly incompressible, anisotropic material with a non-linear
stress-strain relationship. Mechanical deformation was described by Cauchy's
equation of motion under stationary equilibrium assumptions leading to a
quasi-static boundary value problem
\begin{equation}\label{equ:bvp}
  -\nabla_\vec{X}\cdot\tensor{F}\tensor{S}(\vec{d}_\mathrm{s},t) = \vec{0}
  \quad \mbox{in }  {\Omega}_\mathrm{s}^0 ,
\end{equation}
for $t \in [0, T]$, where $\vec{d}_\mathrm{s}$ is the unknown displacement;
$\tensor{F}$ is the deformation gradient;
$\tensor{S}$ is the second Piola--Kirchhoff stress tensor;
and $(\nabla_\vec{X}\;\cdot)$ denotes the divergence operator in the Lagrange
reference configuration. Homogeneous Dirichlet boundary conditions
\begin{equation}\label{eq:dirichlet}
  \vec{d}_\mathrm{s}=\vec{0}\quad \text{on}\quad \Gamma_{\mathrm{s,D}}^0,
\end{equation}
homogeneous Neumann boundary conditions
\begin{equation}\label{eq:neumannH}
    \tensor{F}\tensor{S}(\vec{d}_\mathrm{s},t)\,\vec{n}_\mathrm{s,0}
    = \vec{n}_\mathrm{s,0}
    \quad \text{on} \quad {\Gamma}_{\mathrm{s,H}}^0,
\end{equation}
and inhomogeneous Neumann boundary conditions
\begin{equation}\label{eq:neumann}
  \tensor{F}\tensor{S}(\vec{d}_\mathrm{s},t)\,\vec{n}_\mathrm{s,0}
  = p(t)J\,\tensor{F}^{-\top}(\vec{d}_\mathrm{s},t)\,\vec{n}_\mathrm{s,0}
  \quad \text{on} \quad {\Gamma}_{\mathrm{s,N}}^0
\end{equation}
were imposed, where $\vec{n}_\mathrm{s,0}$ is the outward unit normal vector;
$p(t)$ is the pressure; and $J=\det\tensor{F}$.
For sake of clarity, boundary conditions are illustrated in
\autoref{fig:mechanics_model_generation}~(C).

The total stress $\tensor{S}$ was additively decomposed according to
\begin{equation} \label{eq:additiveSplit}
  \tensor{S}= \tensor{S}_\mathrm{pas}+ \tensor{S}_\mathrm{act},
\end{equation}
where $\tensor{S}_\mathrm{pas}$ and $\tensor{S}_\mathrm{act}$
refer to the passive and active stresses, respectively.
Passive stresses were modeled based on the constitutive equation
\begin{equation}
  \tensor{S}_\mathrm{pas}=2\frac{\partial\Psi(\tensor{C})}{\partial\tensor{C}}
\end{equation}
given a hyper-elastic strain-energy function $\Psi$ and
the right Cauchy--Green strain tensor $\tensor{C}=\tensor{F}^\top\tensor{F}$.
Two different strain-energy functions were used for characterizing passive
mechanical behavior in the LV and the aorta.
In the LV, where the underlying mesh $\Omega_{\mathrm{s,lv}}^0$ and
fiber orientations $(\vec{f}_0, \vec{s}_0,\vec{n}_0)$ are the same
as for the EP model, Section~\ref{sec:eikonal},
the transversely isotropic constitutive relation
\begin{equation} \label{eq:guccioneStrainEnergy}
  \Psi_{\mathrm{Guc}}(\tensor{C}) = \frac{\kappa}{2} \left( \log\,J \right)^2 +
  \frac{C_\mathrm{Guc}}{2}\left[\exp(\mathcal{Q})-1\right].
\end{equation}
by \citet{guccione95} was employed.
Here, the term in the exponent is
\begin{equation} \label{eq:guccioneQ}
  \mathcal{Q} =
    b_{\mathrm{f}} (\vec{f}_0\cdot\isotensor{E}\vec{f}_0)^2 +
    b_{\mathrm{t}} \left[(\vec{s}_0\cdot\isotensor{E}\vec{s}_0)^2+
                         (\vec{n}_0\cdot\isotensor{E}\vec{n}_0)^2+
                        2(\vec{s}_0\cdot\isotensor{E}\vec{n}_0)^2\right]+
    2b_{\mathrm{fs}} \left[(\vec{f}_0\cdot\isotensor{E}\vec{s}_0)^2+
                           (\vec{f}_0\cdot\isotensor{E}\vec{n}_0)^2\right]
\end{equation}
and $\isotensor{E}=\frac{1}{2}(\overline{\tensor C}-\tensor{I})$
is the modified isochoric Green--Lagrange strain tensor, where $\overline{\tensor C} := J^{-\sfrac{2}{3}} \tensor C$.
Default values of $b_\mathrm{f}=18.48$, $b_\mathrm{t}=3.58$,
and $b_\mathrm{fs}=1.627$ were used.
The parameter $C_\mathrm{Guc}$ was varied for the different cases,
see \autoref{tab:_EM_param_fit}.
In the aorta $\Omega_{\mathrm{s,ao}}^0$, unlike in previous studies
\cite{augustin2014classical},
we refrained from assigning fiber structures,
since our efforts were primarily focused on modeling the biomechanics of the LV
and, to a lesser degree, the aorta.
Thus, in absence of information on structural anisotropy, an isotropic model
due to \citet{demiray72:_elasticity} was used
\begin{equation} \label{eq:demirayStrainEnergy}
  \Psi_{\mathrm{Dem}}(\tensor{C}) := \frac{\kappa}{2} \left( \log\,J \right)^2
  + \frac{a}{2\,b} \left\{ \exp \left[ b\,
  \big( \tr(\isotensor{C}) - 3 \big)\right] -1 \right\}.
\end{equation}
The parameter $\widetilde C =\frac{a}{2b}$
was chosen such that $\widetilde C=\SI{3000}{\kPa}$ in the aorta,
$\widetilde C=\SI{30000}{\kPa}$ for valves,
and $\widetilde C=\SI{300}{\kPa}$ for the elastic cushion.
The bulk modulus $\kappa$, which serves as a penalty parameter
to enforce nearly incompressible material behavior, was chosen as
$\kappa = \SI{650}{\kPa}$ in both
Equations~(\ref{eq:guccioneStrainEnergy},~\ref{eq:demirayStrainEnergy}).
For the elastic cushion a value of $\kappa=\SI{100}{\kPa}$ was used.

A simplified phenomenological contractile model was used to represent active
stress generation \cite{niederer11:_length}.
Owing to its small number of parameters and its
direct relation to clinically measurable quantities such as peak pressure,
$p_{\mathrm{lv}}$, and the maximum rate of rise of pressure,
$\dd p_{\mathrm{lv}}/\dt_{\mathrm{max}}$, this model is fairly easy to fit
and thus very suitable for being used in clinical EM modeling studies.
Briefly, the active stress transient is given by
\begin{equation} \label{eq:_tanh_stress}
  S_\mathrm{a}(t,\lambda) = S_\mathrm{peak} \,
  \phi(\lambda) \,
  \tanh^2 \left( \frac{t_\mathrm{s}}{\tau_\mathrm{c}} \right) \,
  \tanh^2 \left( \frac{t_\mathrm{dur} - t_\mathrm{s}}{\tau_\mathrm{r}} \right),
  \qquad \text{for } 0 < t_\mathrm{s} < t_\mathrm{dur},
\end{equation}
with
\begin{equation} \label{equ:_tanh_stress2}
  \phi = \tanh (\mathrm{ld} (\lambda - \lambda_0)),\quad
  \tau_\mathrm{c} = \tau_\mathrm{c_0} + \mathrm{ld}_\mathrm{up}(1-\phi),\quad
  t_\mathrm{s} = t - t_\mathrm{a} - t_\mathrm{emd}
\end{equation}
and $t_\mathrm{s}$ is the onset of contraction;
$\phi (\lambda)$ is a non-linear length-dependent function
in which $\lambda$ is the fiber stretch and
$\lambda_0$ is the lower limit of fiber stretch below which no further active
tension is generated;
$t_{\mathrm{a}}$ is the local activation time from Eq.~\eqref{eq:_eikonal};
$t_{\mathrm{emd}}$ is the EM delay between the onsets of
electrical depolarization and active stress generation;
$S_{\mathrm{peak}}$ is the peak isometric tension;
$t_\mathrm{dur}$ is the duration of active stress transient;
$\tau_{\mathrm{c}}$ is time constant of contraction;
$\tau_{\mathrm{c_0}}$ is the baseline time constant of contraction;
$\mathrm{ld}_{\mathrm{up}}$ is the length-dependence of $\tau_{\mathrm{c}}$;
$\tau_{\mathrm{r}}$ is the time constant of relaxation;
and $\mathrm{ld}$ is the degree of length dependence.
Thus, active stresses in this simplified model are only length-dependent,
but dependence on fiber velocity, $\dot{\lambda}$, is ignored.
Unlinke in previous studies \cite{niederer11:_length} we set the nonlinear length-dependent function $\phi(\lambda) = 1$ for the whole simulation.
The active stress tensor in the reference configuration
$\Omega_{\mathrm{s,lv}}^0$ induced in fiber direction $\vec{f}_0$ is defined as
\begin{equation}\label{eq:act}
  \tensor{S}_{\mathrm{a}} =
    S_{\mathrm{a}} \left(\vec{f}_0\cdot\tensor{C}\vec{f}_0\right)^{-1}
    \vec{f}_0 \otimes \vec{f}_0,
\end{equation}
with $S_a$ defined in Equation~\eqref{eq:_tanh_stress}.
This active stress involves a scaling by $\lambda^2 = \vec{f}_0 \cdot \tensor{C} \vec{f}_0$,
see \cite{pathmanathan2009numerical} for details.
%
%------------------------------------------------------------------------------
\subsubsection{Mechanical and Hemodynamic Afterload Models}
\label{sec:afterload_models}
\begin{figure}[htb]
  \centering
  \includegraphics[width=\textwidth,keepaspectratio]{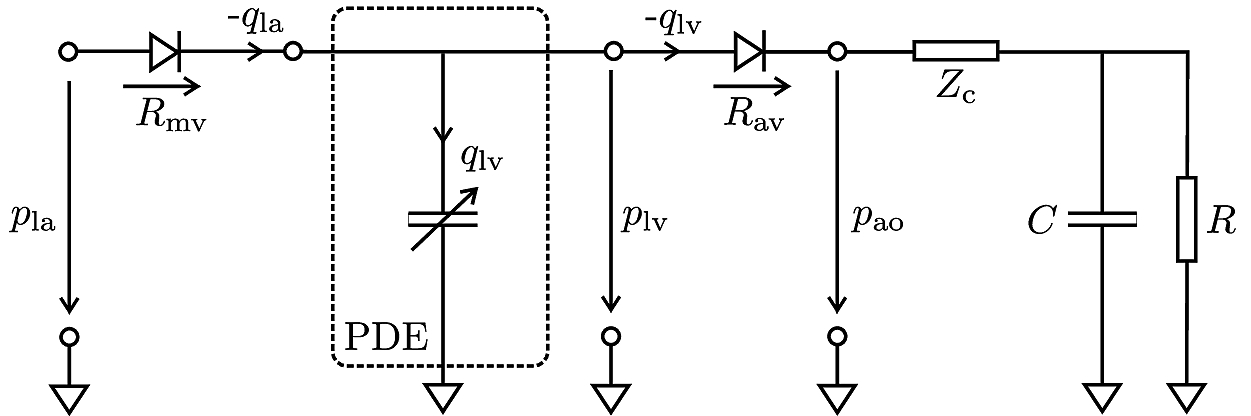}
  \caption{Lumped circuit representation of the coupled EM PDE model of the LV
    with the cardiovascular system.
    The time-varying compliance of the LV is represented as a PDE model
    which was coupled through the aortic valve ($R_\mathrm{av}$) to a 3-element
    Windkessel model
    representing aortic impedance, $Z_\mathrm{c}$, and peripheral arterial
    compliance, $C$, and resistance, $R$, during ejection, and through the
    mitral valve ($R_\mathrm{mv}$) to a constant
    pressure $p_\mathrm{la}$ in the left atrium during filling.
    Negative flows $-q_\mathrm{la}$ and $-q_\mathrm{lv}$ mean the respective
    cavity is ejecting, while positive flow means cavity is being filled.
  }
  \label{fig:_wk3}
\end{figure}

Hydrostatic pressures in the LV, $p_\mathrm{lv}$, and the proximal aorta,
$p_\mathrm{ao}$, were modeled using a 3-element Windkessel model
\cite{westerhof1971artificial}, and the system of PDEs \eqref{equ:bvp} was
linked to this lumped model of the arterial system, see \autoref{fig:_wk3}.
The models were coupled by a diode (aortic valve) which opens at the end of the
isovolumetric contraction (IVC) phase when the pressure in the LV cavity,
$p_\mathrm{lv}$, exceeds the pressure in the proximal aorta, $p_{\mathrm{ao}}$,
and closes at the end of ejection
when $p_\mathrm{lv}$ drops below $p_{\mathrm{ao}}$ and the flow $q_\mathrm{lv}$
starts to reverse. In its open state the aortic valve was modeled as a linear
resistor, $R_\mathrm{av}$, in series with the characteristic impedance of the
aorta, $Z_\mathrm{c}$.
During ejection, the pressure in the LV was then computed by the Windkessel
equation
\begin{equation} \label{equ:_wk3}
  \frac{\dd p_\mathrm{lv}}{\dt} =
    \frac{1}{C} \left(1+\frac{Z_\mathrm{c}+R_\mathrm{av}}{R}\right) q_\mathrm{lv}
    + (Z_\mathrm{c}+R_\mathrm{av}) \frac{\dd q_\mathrm{lv}}{\dt} - \frac{1}{RC} \, p_\mathrm{lv},
\end{equation}
which predicts the rate of change of pressure in the LV as a function of
flow $q_\mathrm{lv}$ out of the LV into the aorta.
The resistor $R$ represents peripheral arterial resistance placed in parallel
with a capacitor $C$, representing vascular compliance.

A similar form of Equation~\eqref{equ:_wk3} was also used to estimate the
pressure in the aorta, $p_\mathrm{ao}$. In this case, there is no additional resistance
due to an outlet valve and hence $R_{\mathrm{av}}$ is omitted.
Balancing of the PDE \eqref{equ:bvp} and the ODE \eqref{equ:_wk3} was achieved
by recasting Equation~\eqref{equ:bvp}
as a saddle point problem, see \cite{gurev15:_high, CNM:CNM2842}.

For CFD simulations, hydrostatic pressures at artificial aortic fluid
outlets, were modeled using a similar 3-element Windkessel model as in
Equation~\eqref{equ:_wk3} that was rewritten in the form of the following
differential algebraic equations for outlet $i$
\begin{align}
  C_i \frac{\mathrm{d}p_{\mathrm d,i}}{\mathrm{d}t}
    + \frac{p_{\mathrm d,i}}{R_i} &= q_i, \label{eq:cfd_wk:1} \\
  p_{\mathrm{wk},i} &= Z_i q_i + p_{\mathrm d,i}, \label{eq:cfd_wk:2}
\end{align}
see \cite{fouchet2014,bertoglio2017} for more details.
During ejection the Windkessel pressure $p_\mathrm{wk}$ at an outlet was then
applied as an outflow boundary condition for the fluid flow model, see
Section~\ref{sec:cfd_boundary_conditions}.
In Equations~(\ref{eq:cfd_wk:1},~\ref{eq:cfd_wk:2}), $C_i$ represents compliance,
$Z_i$ impedence, and $R_i$ resistance of the peripheral arteries for the
respective aortic outlet and $q_i$ denotes the flux through this outlet.
Fitting of the parameters involved will be discussed in
Section~\ref{sec:cfd_boundary_conditions}.
%
%------------------------------------------------------------------------------
\subsection{Fluid flow model} \label{sec:FluidFlowModel}
Human blood in larger vessels such as the LV or the aorta complies with the
assumptions of an incompressible, isothermal, Newtonian and single-phase liquid
\cite{nichols2011mcdonald}.
Let $\Omega_\mathrm{f}\subsetneq \mathbb R^3$ denote the fluid domain, then
the evolution of flow is governed by the incompressible Navier--Stokes
equations
\begin{align}
\rho_\mathrm{f} \left(\partd{}{t}\vec u_{\mathrm{f}} + \vec u_\mathrm{f}\cdot\nabla_{\vec x}\vec u_\mathrm{f}\right)
    - \nabla_{\vec x}
    \cdot \tens \sigma_\mathrm{f}(\vec u_{\mathrm{f}}, p_\mathrm{f})
    &= \vec 0
    && \text{in }\Omega_\mathrm{f}, \label{eq:navier_stokes_pointwise:pde:1}\\
  \nabla_{\vec x} \cdot \vec u_{\mathrm{f}} &= 0
    && \text{in }\Omega_\mathrm{f},\label{eq:navier_stokes_pointwise:pde:2}\\
  \vec u_{\mathrm{f}} &= \vec 0
    && \text{on } \Gamma_\text{noslip},\label{eq:navier_stokes_pointwise:BC:1}\\
  \vec u_{\mathrm{f}} &= \vec g_\mathrm{f}
    && \text{on } \Gamma_\text{inflow},\label{eq:navier_stokes_pointwise:BC:2}\\
  \tens \sigma_\mathrm{f}\vec n_\mathrm{f}
    - \rho_\mathrm{f}\beta\left(\vec u_{\mathrm{f}}
    \cdot \vec n_\mathrm{f}\right)_{-}\vec{u}_{\mathrm{f}}
    &= -p_\mathrm{wk} \vec n_\mathrm{f}
    && \text{on }\Gamma_\text{outflow},\label{eq:navier_stokes_pointwise:BC:3}\\
  \vec u_{\mathrm{f}}\einschraenkung{t=0}
    &= \vec u_0, \label{eq:navier_stokes_pointwise:BC:4}
\end{align}
where $\vec u_{\mathrm{f}}$ denotes fluid velocity;
$p_\mathrm{f}$ is fluid pressure;
$\rho_\mathrm{f}$ is the density of blood;
$\tens \sigma_\mathrm{f}$ is the fluid stress tensor;
$\vec g_\mathrm{f}$ is a velocity inlet;
$p_\mathrm{wk}$ is the Windkessel pressure solution to
Equations~(\ref{eq:cfd_wk:1},~\ref{eq:cfd_wk:2});
$\vec u_0$ refers to the initial condition;
$\vec n_\mathrm{f}$ is the outward normal of the fluid domain;
and $(\nabla_{\vec x})$ is the gradient and $(\nabla_{\vec x}\cdot)$
is the divergence operator in the fluid domain $\Omega_\mathrm{f}$.
The sets $\Gamma_\text{noslip}$, $\Gamma_\text{inflow}$, and $\Gamma_\text{outflow}$
denote the complementary subsets of $\Gamma_\text{f} := \partial \Omega_\text{f}$
and we assume that $\abs{\Gamma_\text{outflow}} > 0$.
Note that Equation~\eqref{eq:navier_stokes_pointwise:BC:2} is given only for the
sake of completeness but was not used in this study, as the inflow of blood
into the aorta is driven by the motion of the LV thus avoiding the need for
prescribing an inflow profile as it is necessary in models
which consider the aorta in isolation.
For $p_\mathrm{wk} \equiv 0$, boundary condition
\eqref{eq:navier_stokes_pointwise:BC:3} is referred to as \emph{directional do-nothing boundary condition}
\cite{EsmailyMoghadam2011, braack2014directional} and the term
\begin{equation}
    \left(\vec u_{\mathrm{f}} \cdot \vec n_\mathrm{f}\right)_{-} :=
    \frac{1}{2}\left(\vec u_{\mathrm{f}}\cdot \vec n_\mathrm{f}
    - \abs{\vec u_{\mathrm{f}} \cdot \vec n_\mathrm{f}}\right)
\end{equation}
is added for backflow stabilization.
A value of $\beta > \frac{1}{2}$ was assumed to guarantee stability of the system.
However, in practical applications values of $\beta \leq \frac{1}{2}$ were also used
without causing numerical issues, see \cite{EsmailyMoghadam2011}.
All physical parameters in Equations~\eqref{eq:navier_stokes_pointwise:pde:1}%
--\eqref{eq:navier_stokes_pointwise:BC:4}
are summarized in \autoref{tab:parameters_navier_stokes}.
In presence of multiple outlets outflow boundary conditions as given in
Equation~\eqref{eq:navier_stokes_pointwise:BC:3} were prescribed at each of the outlets.
%
%------------------------------------------------------------------------------
\subsubsection{Extension to Moving Geometries}
For time-dependent fluid domains, i.e., $\Omega_\text{f} = \Omega_\text{f}^t$,
Equations~\eqref{eq:navier_stokes_pointwise:pde:1}%
--\eqref{eq:navier_stokes_pointwise:BC:4}
need to be modified to account for the domain movement.
This requires the linking of the equations governing fluid dynamics -- posed in
an Eulerian coordinate frame -- with the structural mechanics equations --
posed in a Lagrangian reference frame. This is achieved by using the ALE
formulation which combines both
Lagrangian and Eulerian formulation in a generalized description,
see \cite[Section 1.3]{bazilevs2013computational} and \cite{hirt1974arbitrary}.
Similar to structural mechanics, a reference fluid configuration
$\Omega_\text{f}^0 \subsetneq \mathbb R^3$ is used which we identify
with the mesh been generated at end-diastolic state, see
Section~\ref{sec:anatomic_model_generation}.
The coordinate system of the Eulerian frame is denoted by $\vec x$
and the reference coordinate system is denoted by $\vec X$.
Their relation is given by the ALE mapping
$\vec x = \vec X + \vec d_\mathrm{f}(t, \vec X)$.
Here, $\vec d_\mathrm{f}(t,\vec X)$ refers to an arbitrary, not necessarily
physical, displacement of points to track the deformation of the fluid domain.
Using this ALE mapping the time-dependent moving fluid domain is represented as
\begin{equation} \label{eq:def_omega_cfd}
  \Omega_\text{f}^t := \left\{\vec x : \vec x = \vec X
    + \vec d_\mathrm{f}(t,\vec X),\,
  \forall \vec X \in \Omega_\text{f}^0\right\}.
\end{equation}
Further, we define the fluid domain velocity $\vec w_\mathrm{f}$ as
\begin{equation}
  \vec w_\mathrm{f} := \partd{}{t} \vec d_\mathrm{f}\einschraenkung{\vec X},
\end{equation}
where $\partd{}{t}(\cdot)\einschraenkung{\vec X}$ is the derivative with
respect to $t$ with $\vec X$ being fixed,
and the moving interface between fluid and solid domain as
\begin{equation}
    \Gamma_{\mathrm{f,mov}}^t := \partial \Omega_\text{f}^t \  \backslash
    \bigcup_{i=1}^{n_\text{outlets}}\Gamma_{\mathrm{f,outflow},i}^t,
\end{equation}
where $\Gamma_{\mathrm{f,outflow},i}^t$ are the individual aortic outlets.
The fluid displacement at this point remains unknown and will be specified in
Section~\ref{sec:transfer}.
Combining these concepts, an ALE description of the Navier--Stokes equations can
be derived, see, e.g., \cite{bazilevs2013computational,forster2006geometric},
\begin{align}
  \rho_\mathrm{f} \left(\partd{}{t}\vec{u}_{\mathrm{f}}\einschraenkung{\vec X}
    + \left(\vec u_\mathrm{f}-\vec w_\mathrm{f}\right)\cdot\nabla_{\vec x}\vec u_\mathrm{f}\right)
    - \nabla_{\vec x}
    \cdot \tens \sigma_\mathrm{f}(\vec u_{\mathrm{f}}, p_{\mathrm f}) &= \vec 0
    && \text{on }\Omega_\mathrm{f}^t, \label{eq:ale_pointwise:pde:1}\\
  \nabla_{\vec x} \cdot \vec u_\mathrm{f} &= 0
    && \text{on } \Omega_\mathrm{f}^t,\label{eq:ale_pointwise:pde:2}\\
  \vec u_\mathrm{f} &= \vec{g}_\mathrm{mov}
    && \text{on } \Gamma_{\mathrm{f,mov}}^t,\label{eq:mov_dom_bc}\\
  \tensor{\sigma}_\mathrm{f}(\vec u_{\mathrm{f}}, p_\mathrm{f})\vec{n}_\mathrm{f}
    - \rho_\mathrm{f} \beta ((\vec u_{\mathrm{f}}-\vec w_{\mathrm{f}})
    \cdot \vec n_\mathrm{f})_{-}\vec u_{\mathrm{f}}
    &= -p_\mathrm{wk,i}\vec n_\mathrm{f} && \text{on each }
    \Gamma_{\mathrm{f,outflow},i}^t,\label{eq:outflow_stabil:ale}\\
  \vec u_{\mathrm{f}}\einschraenkung{t=0} &= \vec u_0
    && \text{in } \Omega_\mathrm{f}^0. \label{eq:ale:initial}
\end{align}
%
%Here $\Gamma_\text{mov}$ denotes the subset of
%$\Gamma(t) := \partial \Omega_\mathrm{f}(t)$ which undergoes a movement.
Along $\Gamma_{\mathrm{f,mov}}^t$ we imposed equality between fluid velocity and
the velocity of the moving surfaces.
Boundary condition \eqref{eq:outflow_stabil:ale} is the ALE equivalent of the
outflow stabilization in Equation~\eqref{eq:navier_stokes_pointwise:BC:3},
see \cite[Section 8.4.2.3]{bazilevs2013computational}. Details on how domain
movement and velocity were chosen in our application will be discussed later in
Sections~\ref{sec:transfer} and \ref{sec:cfd_boundary_conditions}.
%
%------------------------------------------------------------------------------
\subsubsection{Variational Formulation of the Navier--Stokes equations}
Following \cite{bazilevs2007variational,bazilevs2013computational,pauli2017},
the discrete variational formulation of the ALE equations
\eqref{eq:ale_pointwise:pde:1}--\eqref{eq:ale:initial}
can be stated in the following abstract form:
find $\vec{u}^\mathrm{h}_\mathrm{f} \in
[\mathcal{S}_{\mathrm{h},\vec g}^1(\mathcal T_\mathrm{N})]^3,
p^\mathrm{h}_\mathrm{f} \in \mathcal S_\mathrm{h}^1(\mathcal T_\mathrm{N})$
such that for all
$\vec v^\mathrm{h} \in [\mathcal S_{h,\vec 0}^1(\mathcal T_\mathrm{N})]^3$
and for all $q^\mathrm{h} \in \mathcal S_\mathrm{h}^1(\mathcal T_\mathrm{N})$
\begin{equation} \label{eq:ale_vms_discrete}
  A_\mathrm{NS}(\vec v^\mathrm{h},q^\mathrm{h};
    \vec u^\mathrm{h}_\mathrm{f},p^\mathrm{h}_\mathrm{f})
  + S_\mathrm{VMS}(\vec v^\mathrm{h},q^\mathrm{h};
    \vec u^\mathrm{h}_\mathrm{f}, p^\mathrm{h}_\mathrm{f})
  = F_\mathrm{NS}(\vec v^\mathrm{h}),
\end{equation}
with the classical bilinear form of the Navier--Stokes equations
\begin{align}
  A_\mathrm{NS}(\vec v^\mathrm{h},q^\mathrm{h};
    \vec u^\mathrm{h}_\mathrm{f},p^\mathrm{h}_\mathrm{f})
    := &\rho_\mathrm{f}\int\limits_{\Omega_\mathrm{f}^t}\vec v^\mathrm{h}
    \cdot \left(\partd{}{t}\vec u^\mathrm{h}_\mathrm{f}
    + \left(\vec u_\mathrm{f}^\mathrm{h}-\vec w_\mathrm{f}^\mathrm{h}\right)\cdot\nabla_{\vec x}\vec u_\mathrm{f}^\mathrm{h}\right)\dx
    + \int\limits_{\Omega_\mathrm{f}^t}\tens\varepsilon(\vec v^\mathrm{h})
    :\tensor{\sigma}_\mathrm{f}(\vec u^\mathrm{h}_\mathrm{f},
    p^\mathrm{h}_\mathrm{f})\dx \nonumber\\
  & + \int\limits_{\Omega_\mathrm{f}^t} q^\mathrm{h} \nabla_{\vec x}
    \cdot \vec u^\mathrm{h}_\mathrm{f} \dx
    - \rho_\mathrm{f}\beta\sum_{i=1}^{n_\text{outlets}}
    \int\limits_{\Gamma_{\mathrm{f,outflow},i}^t} ((\vec u^\mathrm{h}_\mathrm{f}
    - \vec w^\mathrm{h}_\mathrm{f})\cdot\vec n_\mathrm{f})_{-} \vec v^\mathrm{h}
    \cdot \vec u^\mathrm{h}_\mathrm{f}\dsx, \label{eq:ale_ANS}
\end{align}
the bilinear form $S_\mathrm{VMS}$, which is explained later in
Equation~\eqref{eq:ale:Svms}, and the right-hand side contribution
\begin{equation}
  F_\text{NS}(\vec v^\mathrm{h}) := -\sum_{i=1}^{n_\text{outlets}}
    p_\mathrm{wk,i}\int\limits_{\Gamma_{\mathrm{f,outflow},i}^t}  \vec v^\mathrm{h}
    \cdot \vec n_\mathrm{f}\dsx.
\end{equation}
In Equation~\eqref{eq:ale_ANS}, $\tensor{\varepsilon}$ is the strain-rate tensor and
$\vec w^\mathrm{h}_\mathrm{f}$ is the discrete counterpart of the
fluid domain velocity $\vec w_\mathrm{f}$, i.e.,
\begin{equation}
  \vec w^\mathrm{h}_\mathrm{f}(t^{n+1},\vec X)
    = \frac{\vec d_\mathrm{f}(t^{n+1},\vec X)
    - \vec d_\mathrm{f}(t^{n},\vec X)}{\Delta t}.
\end{equation}
The FE function space
$\mathcal S_{\mathrm{h},*}^1(\mathcal T_\mathrm{N})$ is the conformal trial
space of piecewise linear, globally continuous basis functions over a
decomposition $\mathcal T_\mathrm{N}$ of $\Omega_\mathrm{f}^t$ into
$N$ simplicial elements constrained by $\vec v^\mathrm{h} = \ast$
on essential boundaries.
The FE function space
$\mathcal S_\mathrm{h}^1(\mathcal T_\mathrm{N})$ denotes the same space
without constraints. For further details we refer to
\cite{brenner2007mathematical,steinbach2007numerical}.

From a mathematical point of view, the Navier--Stokes equation can be seen as a
multidimensional convection–diffusion equation with pressure acting as a
Lagrangian multiplier of the incompressibility constraint.
In the common case where velocity and pressure are retained as unknowns, as above,
the Ladyzhenskaya--Babu\v{s}ka--Brezzi (LBB) condition has to be satisfied by
the velocity and pressure spaces \cite{donea2003finite}.
A violation of the LBB condition may lead to pressure oscillations.
Stabilization techniques allowing the circumvention of the LBB condition exist
and have been extensively studied, see for example
\cite{hughes1986new,franca1988two,douglas1989absolutely,bochev2006stabilization}.
However, with increasing Reynolds number the Navier--Stokes equations become convection
dominated. This requires increasingly finer mesh resolutions to accurately
resolve finer flow details which, eventually, renders numerical solution in
this form computationally intractable.
As a remedy, one can resort to using turbulence models.
In particular, in this study the
\emph{residual based variational multiscale turbulence model} (RBVMS), see
\cite{hughes1995multiscale, bazilevs2007variational,bazilevs2013computational,%
pauli2017} was employed which acts as a stabilization and a turbulence model.
The underlying main idea is to split the unknown solution into
resolvable (coarse) and unresolvable (fine) scales by the FE
approximation, where the finer scale details are taken into account based
on element residuals. For details on the derivation we refer to elsewhere
\cite{bazilevs2007variational}.
The term $S_\mathrm{VMS}$ in Equation~\eqref{eq:ale_vms_discrete} denotes the
bilinear form of the RBVMS formulation and reads as
\begin{align}
  S_\mathrm{VMS}(\vec v^\mathrm{h},q^\mathrm{h};
    \vec u^\mathrm{h}_\mathrm{f}, p^\mathrm{h}_\mathrm{f})
  := &\frac{1}{\rho_\mathrm{f}}\sum\limits_{l=1}^{n_\text{el}}
    \int\limits_{\tau_\ell}\tau_\text{MOM}
    \left(\rho_\mathrm{f}\left(\vec u^\mathrm{h}_\mathrm{f} - \vec w^\mathrm{h}_\mathrm{f}\right)\cdot\nabla_{\vec x} \vec v^\mathrm{h}
    + q^\mathrm{h}\right)
    \cdot\vec r_\text{MOM}(\vec u^\mathrm{h}_\mathrm{f},
    p^\mathrm{h}_\mathrm{f})\dx \nonumber\\
  & + \sum\limits_{l=1}^{n_\text{el}}\int\limits_{\tau_\ell}
    \tau_\text{CONT}\nabla_{\vec x} \cdot \vec v^\mathrm{h} \nabla_{\vec x}
    \cdot \vec u^\mathrm{h}_\mathrm{f}\dx \nonumber\\
  & - \sum\limits_{l=1}^{n_\text{el}}\int\limits_{\tau_\ell}
    \tau_\text{MOM}\vec v^\mathrm{h}
    \cdot \left(\nabla_{\vec x} \vec u^\mathrm{h}_\mathrm{f}
    \vec r_\text{MOM}(\vec u^\mathrm{h}_\mathrm{f},
    p^\mathrm{h}_\mathrm{f})\right)\dx \nonumber \\
  & - \frac{1}{\rho_\mathrm{f}}\sum\limits_{l=1}^{n_\text{el}}
    \int\limits_{\tau_\ell}
    \tau_\text{MOM}^2\tens\varepsilon(\vec v^\mathrm{h})
    : (\vec r_\text{MOM}(\vec u^\mathrm{h}_\mathrm{f}, p^\mathrm{h}_\mathrm{f})
    \otimes \vec r_\text{MOM}(\vec u^\mathrm{h}_\mathrm{f},
    p^\mathrm{h}_\mathrm{f}))\dx,
    \label{eq:ale:Svms}
\end{align}
where the vector $\vec r_\text{MOM}$ is defined as
\begin{equation}
  \vec r_\text{MOM}(\vec u^\mathrm{h}_\mathrm{f}, p^\mathrm{h}_\mathrm{f})
    := \rho_\mathrm{f} \left(\partd{}{t}\vec u^\mathrm{h}_\mathrm{f}
      + \left(\vec u^\mathrm{h}_\mathrm{f} - \vec w^\mathrm{h}_\mathrm{f}\right)\cdot\nabla_{\vec x}\vec u^\mathrm{h}_\mathrm{f}\right)
      - \nabla_{\vec x}
      \cdot \tens \sigma_\mathrm{f}(\vec u^\mathrm{h}_\mathrm{f}, p^\mathrm{h}_\mathrm{f}).
\end{equation}
The definition of the parameters $\tau_\text{MOM}, \tau_\text{CONT}$
according to \cite{pauli2017} is given by
\begin{equation}
  \tau_\text{MOM} := \min\left\{\left(\frac{4}{\Delta t^2}
    + (\vec u^\mathrm{h}_\mathrm{f}-\vec w^\mathrm{h}_\mathrm{f})
    \cdot \tens G(\vec u^\mathrm{h}_\mathrm{f} - \vec w^\mathrm{h}_\mathrm{f})\right)^{-\frac{1}{2}},
    \frac{\rho_\mathrm{f} C_\mathrm{M}}
         {\mu_\mathrm{f} \sqrt{\tens G : \tens G}}\right\},
\end{equation}
with $\Delta t$ being the time step size and
$\tens G := \partd{\vec \xi}{\vec x}^\top \tens K \partd{\vec \xi}{\vec x}$,
where $\partd{\vec \xi}{\vec x}$ denotes the Jacobian of the mapping from a
physical FE to the reference FE,
the tensor $\tens K$ is defined as
\begin{equation}
  \tens K := \frac{1}{2\sqrt[3]{2}}
    \begin{pmatrix}
      3 & -1 & -1 \\
     -1 &  3 & -1 \\
     -1 & -1 &  3
    \end{pmatrix}
\end{equation}
and the constant $C_\mathrm{M} = 0.0285$.
Further, the stabilization parameter $\tau_\text{CONT}$ is defined as
\begin{align}
  \tau_\text{CONT} &:= \frac{1}{\tau_\text{MOM} \vec{g}_\mathrm{f} \cdot \vec{g}_\mathrm{f} },\\
  g_{\mathrm{f},i} &:= \sum\limits_{j=1}^3 \left(\partd{\vec \xi}{\vec x}\right)_{ji}.
\end{align}
%
%------------------------------------------------------------------------------
\subsubsection{EM-based Kinematic driver model}
\label{sec:transfer}
Displacements computed with the EM model were used to prescribe the kinematics
of the blood pool mesh which in turn was used for simulating hemodynamics
in the CFD model.
This was achieved by imposing
$\vec{g}_\mathrm{mov} = \partd{}{t}\vec{d}_\mathrm{s}$ in Equation~\eqref{eq:mov_dom_bc}.
Since the surface of the reference CFD blood pool mesh,
$\partial \Omega_\mathrm{f}^0$, is not conformal with the surface of the
reference EM blood pool mesh, $\Omega_{\mathrm{s,bp}}^0$, and the overlap of
the two surfaces is imperfect due to smoothing of
$\partial \Omega_\mathrm{f}^0$ and remeshing of $\Omega_\mathrm{f}^0$,
a direct transfer of displacements between the two surfaces is not
readily feasible. As a remedy, we proceeded as follows.
After solving the EM problem the subset of displacements
$\widetilde{\vec{d}}_\mathrm{s}$ that form the endocardial interface with
the blood pool, $\Gamma_{\mathrm{s,bp}}^0$, were extracted from the
solution $\vec d_\mathrm{s}$ defined at $\Omega_\mathrm{s}^0$.
Since the mesh interface between $\Omega_\mathrm{s}^0$ and
$\Omega_{\mathrm{s,bp}}^0$ is conformal the extracted displacements can be
applied as inhomogeneous time-varying Dirichlet boundary conditions to the
blood pool mesh $\Omega_{\mathrm{s,bp}}^0$ to solve a
linear elastic problem given as
    \begin{align}
      -\nabla_{\vec X} \cdot \tens\sigma(\vec d_\mathrm{s}(t)) &= \vec 0 &&
        \text{in } \Omega_{s,\mathrm{bp}}^0,
        \label{eq:deform_EM_blood_pool:1}\\
      \vec d_\mathrm{s}(t) &= \widetilde{\vec{d}}_\mathrm{s}(t) &&
        \text{on } \partial \Omega_{s,\mathrm{bp}}^0,
    \end{align}
where stress and strain tensor are
  \begin{align}
    \tens \sigma(\vec d_\mathrm{s}) &:=
      \frac{E}{1+\nu}\left(\frac{\nu}{1-2\nu}\nabla_{\vec X}
      \cdot \vec d_\mathrm{s} \tens I
      + \tensor{\varepsilon}(\vec d_\mathrm{s}) \right), \\
    \tensor{\varepsilon}(\vec d_\mathrm{s}) &:= \frac{1}{2}\left(
      \nabla_{\vec X} \vec d_\mathrm{s}
      + (\nabla_{\vec X} \vec d_\mathrm{s})^\top\right),
      \label{eq:deform_EM_blood_pool:4}
  \end{align}
the constant $E$ is Young's modulus in $\si{\kilo\pascal}$ and the constant $\nu$ is
Poisson's ratio which is dimensionless in the range of $[-1,0.5)$.
Combining the solutions $\vec d_\mathrm{s}$ computed for $\Omega_\mathrm{s}^0$
and $\Omega_{\mathrm{s,bp}}^0$ yields displacements $\vec d_\mathrm{s}$
for $\Omega_{\mathrm{s,total}}^0$.
Since $\partial \Omega_\mathrm{f}^0$ is fully embedded in this domain,
$\Omega_{\mathrm{s,total}}^0$ $\Omega_{\mathrm{s,total}}^0$ can be used as a
hanging background mesh for interpolating displacements onto the blood pool
mesh, $\Omega_\mathrm{f}^0$, used for CFD simulations.
However, for reasons of mesh quality, interpolation is solely applied on the boundary
$\Omega_\mathrm{f}^0$ itself, and to find the interior displacement field the exact
same linear elastic problem
\eqref{eq:deform_EM_blood_pool:1}--\eqref{eq:deform_EM_blood_pool:4},
is solved for $\vec{d}_\mathrm{f}$ instead of $\vec{d}_\mathrm{s}$.
In both patient cases studied, ejection fractions were large leading to a
substantial deformation of the blood pool mesh $\Omega_\mathrm{f}^t$.
To maintain mesh quality under such large deformations
the parameters $E$ and $\nu$ governing stiffness and incompressibility of the
material were altered accordingly.
Initially, a fixed $E_0$ and $\nu_0$ was chosen while the subsequent
modification of $E$ and $\nu$ was guided by a combination of the two following
strategies.
\begin{enumerate}[i) ]
  \item \emph{Quality based stiffening:}
    For each element $\tau_\ell$ in the fluid mesh a tetrahedral quality
    indicator $\kappa(\tau_\ell)$ based on the movement from the previous time
    step was calculated, see \cite{freitag2002tetrahedral,kanchi20073d},
    and rescaled such that for elements of good quality $\kappa$ is close to
    1, while for elements with poor quality $\kappa$ tends towards infinity.
    Eventually, the parameter $E$ was multiplied by $\kappa$ within each element.
  \item \emph{$\nu$-Volume based stiffening:} For larger deformation
    elements in the fluid mesh may collapse or even invert, yielding a zero or
    negative volume. When solving
    Equations~\eqref{eq:deform_EM_blood_pool:1}--\eqref{eq:deform_EM_blood_pool:4},
    the current element volumes were tracked and a volume ratio relative to an
    undeformed reference element was computed as
    $\frac{\abs{\tau_\ell}}{\abs{\hat{\tau}_\ell}}$.
    For ratios below a predefined critical value the parameter $\nu$ was
    set close to $0.5$ to make this element nearly incompressible.
\end{enumerate}
%
%------------------------------------------------------------------------------
\subsection{Numerical Solution} \label{sec:NumericalSolution}
Spatio-temporal discretization of all PDEs and the solution of the arising
systems of equations relied upon the Cardiac Arrhythmia Research Package
(CARP), see \citet{vigmond2003carp}.
Numerical details on FE discretization \cite{rocha2011hybrid}
and solution of EP
\cite{vigmond08:_solvers,neic12:_accelerating,neic17:_efficient}
and EM \cite{augustin16:_anatomically}
have been discussed in detail elsewhere.
FE discretization and solution of the Navier--Stokes equations were implemented
recently using the same numerical framework which was extended to account for
non-linear saddle-point problems arising from the discretized CFD equations.

Two time discretization schemes were implemented and compared for the applications
in mind, and a computationally cheap semi-implicit scheme,
modified from \cite[Section 1.4.2]{forti2016phd}, showed similar
results to the more expensive fully-implicit generalized-$\alpha$ method
\cite{jansen2000generalized}.
Hence, all results in Section~\ref{sec:res} were obtained using the
semi-implicit scheme; to advance from time step $t^n$ to $t^{n+1}$, only a linear
block system needs to be solved, where each block depends on data from the
previous time step only. Solvers for the block system were taken from the PETSc
library \cite{petsc-web-page,petsc-user-ref,petsc-efficient}.
We used a right preconditoned flexible GMRES method with PETSc fieldsplit
preconditioning \cite{silvester2001,elman2008} which in turn uses BoomerAMG
\cite{yang2002boomeramg} to approximate sub-block inverses.
While the time step size for mechanics and CFD was the same,
$\Delta t_\mathrm{mech}=\Delta t_\mathrm{CFD}=\SI{0.5}{\milli\second}$, it was
significantly smaller for EP, where $\Delta t_\mathrm{EP}=\SI{25}{\us}$.

The implementation of the CFD solvers has been subjected to various validation
procedures against standard CFD benchmarks \cite{schafer1996benchmark}.
All simulations were executed at the national HPC computing facility ARCHER
in the United Kingdom using \num{384} and \num{768} cores for EM and CFD
simulations, respectively.
%
%------------------------------------------------------------------------------
\subsection{Model parameterization}
\label{sec:meth:parameterization}
\subsubsection{Electrophysiology}
\label{sec:meth:param_EP}

% EP
Electrical activation sequences were indirectly parameterized
using the QRS complex of a given patient's ECG as guidance.
Unlike in previous studies \cite{augustin16:_patient},
we refrained from a detailed parameterization
which aimed at reproducing the QRS complex of the ECG for a given patient
by finding appropriate locations and timings for the main fascicles
of the cardiac conduction system in the LV.
Rather, default locations and timings were used
which yielded a total activation time within the physiological range.
%
%------------------------------------------------------------------------------
\subsubsection{Passive biomechanics} \label{sec:passive_mech_para}
The LV myocardium was characterized as a hyperelastic, nearly incompressible,
transversely isotropic material with a nonlinear stress–strain relationship
\cite{guccione95}.
Orthotropic material axes were aligned
with the local fiber, sheet and sheet normal directions.
To remove rigid body motion, homogeneous displacement boundary conditions
were applied by fixing the terminal rims of the clipped
brachiocephalic, left common carotid and left subclavian arteries
as well as the clipped rim of the aorta descendens,
see \autoref{fig:mechanics_model_generation}.
The model was stabilized by resting the LV apex on an elastic cushion
of which the bottom face was rigidly anchored
also by applying homogeneous displacement boundary conditions.

% passive biomechanics
The constitutive model was fitted to recorded clinical data
as previously reported with minor modifications \cite{augustin16:_patient}.
The passive biomechanical model governed by the strain-energy function
given in Equation~\eqref{eq:guccioneStrainEnergy}
was fitted to approximate the end-diastolic pressure-volume relation (EDPVR).
Due to limitations in the recorded data we refrained from directly fitting
the model to the recorded pressure and volume data.
Rather, only one data pair
-- EDV and end-diastolic pressure (EDP) -- was used
to fit the stress-free residual volume to the empiric Klotz relation
\cite{klotz2007computational} by adjusting the isotropic scaling parameter
$C_\mathrm{Guc}$
in Equation~\eqref{eq:guccioneStrainEnergy}.
As the model anatomy was built from a segmented 3DWH MRI scan -- acquired during
diastasis -- the FE model was inflated to increase the volume of the cavity by
the difference between the volume at mid diastasis and the EDV.
Using the end-diastolic geometry, default material parameters and the recorded EDP,
an initial guess of the stress-free reference configuration was computed
by unloading the model using a backward displacement method
\cite{sellier2011iterative,bols2013computational,krishnamurthy2013patient}. The unloading procedure was repeated with
varying trial material parameters, $C_\mathrm{Guc}$,
until the difference between the unstressed LV volume of the model
and the prediction of the Klotz relation was less than \SI{5}{\percent}.
%
%------------------------------------------------------------------------------
\subsubsection{Active stresses}
% active stress
Parameters of the active stress model were fitted
during IVC and ejection phase.
During IVC the LV volume was held constant \cite{gurev15:_high} and the
parameters of the active stress given in Equation~\eqref{eq:_tanh_stress} rate of
contraction, $\tau_\mathrm{c}$, and peak active stress, $S_\mathrm{peak}$,
were manually adjusted to fit the maximum rate of rise of pressure,
$(\mathrm{d}P/\mathrm{d}t)_\mathrm{max}$,
and peak pressure, $p_{\mathrm{lv}}$.
%
%------------------------------------------------------------------------------
\subsubsection{Afterload}
% afterload EM model
When the LV pressure $p_{\mathrm{lv}}$ exceeded the aortic pressure,
$p_{\mathrm{ao}}$, ejection was initiated by connecting the LV model with the
lumped 3-element Windkessel model  \cite{westerhof1971artificial}.
Volume traces recorded from a given patient during ejection were used as input
to compute aortic pressure traces by solving Equation~\eqref{equ:_wk3}.
Both types of data were not recorded simultaneously
as volume traces were computed from Cine MRI scans
and pressure traces were recorded later invasively by catheterization.
Volume and pressure traces were synchronized in time
by aligning the onset of ejection of the volume trace $V_{\mathrm{lv}}(t)$
with the instant of opening of the aortic valve in the pressure trace
$p_{\mathrm{ao}}(t)$.
In those cases where heart rates were markedly different between the two
measurements, volume traces were scaled in time to adjust
LV ejection time (LVET) to the duration of ejection in the pressure traces,
that is, the time elapsed between opening and closing of the aortic valve
as these two instants in time were clearly identifiable in all traces
$p_{\mathrm{ao}}(t)$, see \autoref{fig:_results_EM_validation}.
Moreover, volume traces were offset to ensure
that the model volume based on the segmentation of the 3DWH scan acquired
during diastasis matched up with the Cine-MRI based volume trace at mid diastasis.
The parameter space of the Windkessel model comprising characteristic
impedance of the aorta, $Z_\mathrm{c}$, as well as resistance, $R$,
and compliance, $C$, of the arterial system was sampled using a recently
developed stochastic sampling approach \cite{crozier2016relative}.

Numerous box constraints were used to constrain the search space of parameter
sweeps. In particular, we used reported measurements in humans to define the
mean values and restricted the search space for each parameter to fall within
\SI{\pm 20}{\percent} around the mean.
Due to high frequency errors introduced by the pressure transducer we
refrained from computing norms $||p_{\mathrm{ao,meas}} - p_{\mathrm{ao,fit}}||$
to quantify the deviations of fitted from measured pressure
and opted for manual selection using three criteria,
aortic peak pressure, $p_{\mathrm{ao}}$, closing pressure of aortic
valve and exponential decay of $p_{\mathrm{ao}}$ during diastole.
For the sake of fitting $Z_\mathrm{c}$ we assumed
$p_{\mathrm{ao}} \approx p_{\mathrm{lv}}$ since transvalvular pressure
gradients in all patients were very minor.
%
%------------------------------------------------------------------------------
\subsubsection{CFD boundary conditions}
\label{sec:cfd_boundary_conditions}
The validated EM models yield the time-dependent displacement fields,
$\vec d_\mathrm{s}$, which were transferred onto the fluid domain
to drive simulations of blood flow in LV and aorta as described in
Section~\ref{sec:transfer} yielding $\vec d_\mathrm{f}(t,\vec x)$ defined on
the whole CFD mesh. \autoref{fig:fluid_model_generation} \textsf{(G)} shows a summary
of the boundary conditions. On the boundary $\Gamma_{\mathrm f,\mathrm{mov}}^t$ a
Dirichlet boundary condition enforcing the mesh velocity
$\vec w^{\mathrm{h}}_{\mathrm{f}}$ is applied.
On each aortic outlet $\Gamma_{\mathrm{f,outflow},i}(t)$ a 3-Element Windkessel
model as described in Section~\ref{sec:afterload_models} is attached.
Further, the stabilization parameter $\beta$ in
Equation~\eqref{eq:outflow_stabil:ale} was set to \num{0.2}.
Estimation of the input parameters for the hemodynamical Windkessel equations
relied on an extension of the simple hydraulic analog of Ohm's law.
Given the patient specific MAP, CO, and a percentage $\alpha_i$ of total CO
running through the outlet the resistance $R_i$ was estimated as
\begin{equation}
  R_i \approx \frac{\mathrm{MAP}}{\alpha_i \mathrm{CO}}.
\end{equation}
The percentages $\alpha_i$ were obtained either by measurement or by applying
Murray's law \cite{murray1926physiological}.
The impedances $Z_i$ were chosen as
\SI{5}{\percent} of $R_i$, and the compliances $C_i$ were chosen such that
$R_i C_i \approx \SI{1000}{\milli\second}$. To keep the semi-implicit character
of the CFD system the Windkessel equations were solved with a
semi-implicit backward Euler method using the flow $q_i^n$ through the
aortic outlet, from the previous time step as input.
%
%==============================================================================
\section{Results} \label{sec:res}
\subsection{Building electromechanical kinematic driver models}
\label{sec:res:EM}
\begin{figure}[htb]
  \centering
  \includegraphics[height=9cm,keepaspectratio]{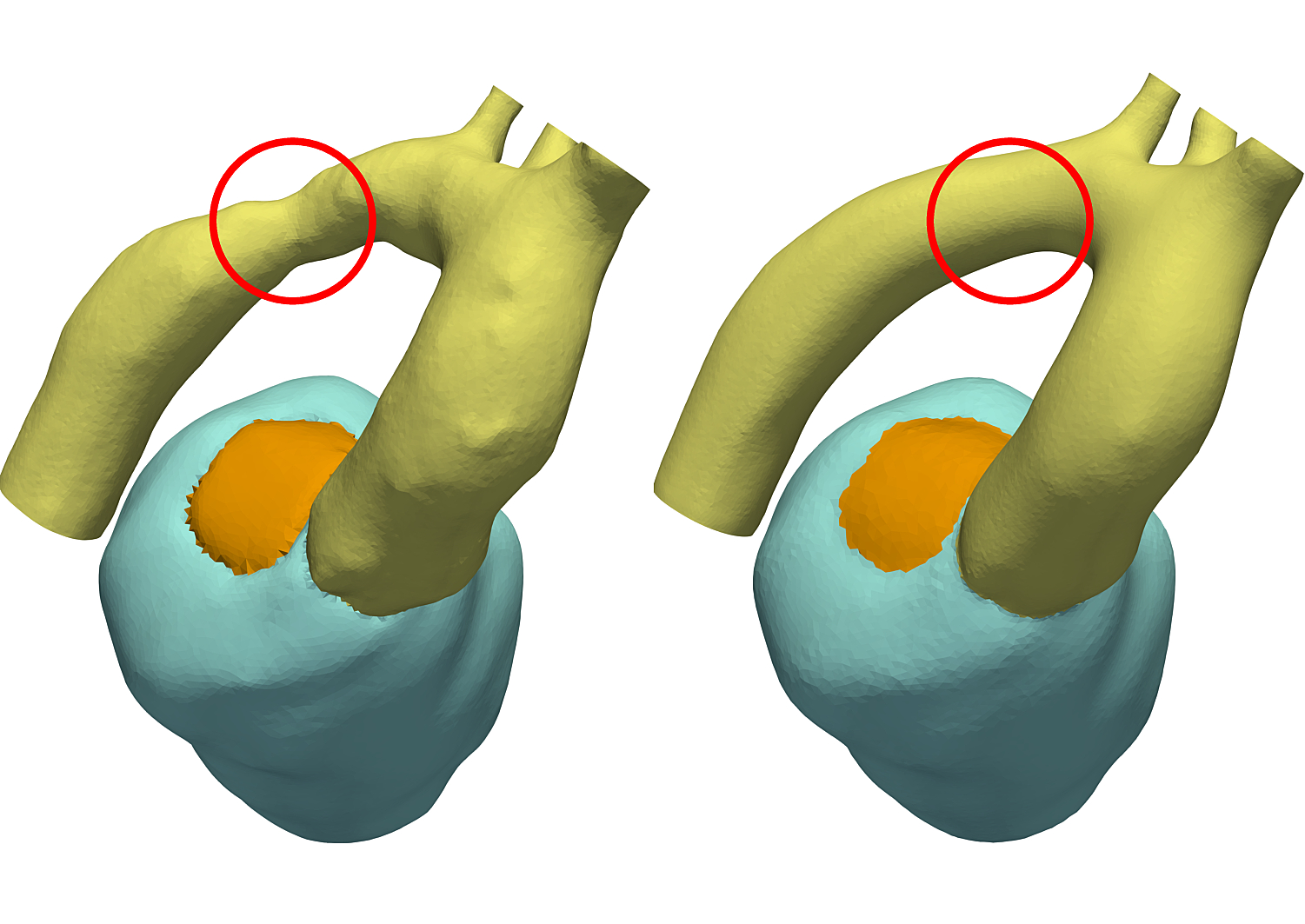}
  \caption{CoA anatomy of case 28 before and after virtual stenting procedure.
  	CoA location is indicated with  a red circle.}
  \label{fig:post_treatment_geo_comp}
\end{figure}

Using a previously developed automated workflow \cite{crozier2016imagebased},
anatomical FE models of LV and aorta were built for patient cases \emph{28-Pre} and \emph{44-Pre}
based on segmented imaging data acquired under pre-treatment conditions.
\autoref{fig:mechanics_model_generation} illustrates the key processing steps and the resulting FE model
for case \emph{28-Pre}.
For the case \emph{28-Pre} the CoA was repaired by a virtual dilatation procedure
applied to the segmented image data with the aim to restore normal cross sectional areas.
Subsequently, a new FE mesh was generated referred to as \emph{28-Post},
which was essentially identical to \emph{28-Pre},
with the only difference being the anatomical adjustment of the CoA in the
aortic arch to the target post-treatment anatomy after stenting, see \autoref{fig:post_treatment_geo_comp}.

\begin{figure}[htb]
  \centering
  \includegraphics[width=\textwidth,keepaspectratio]{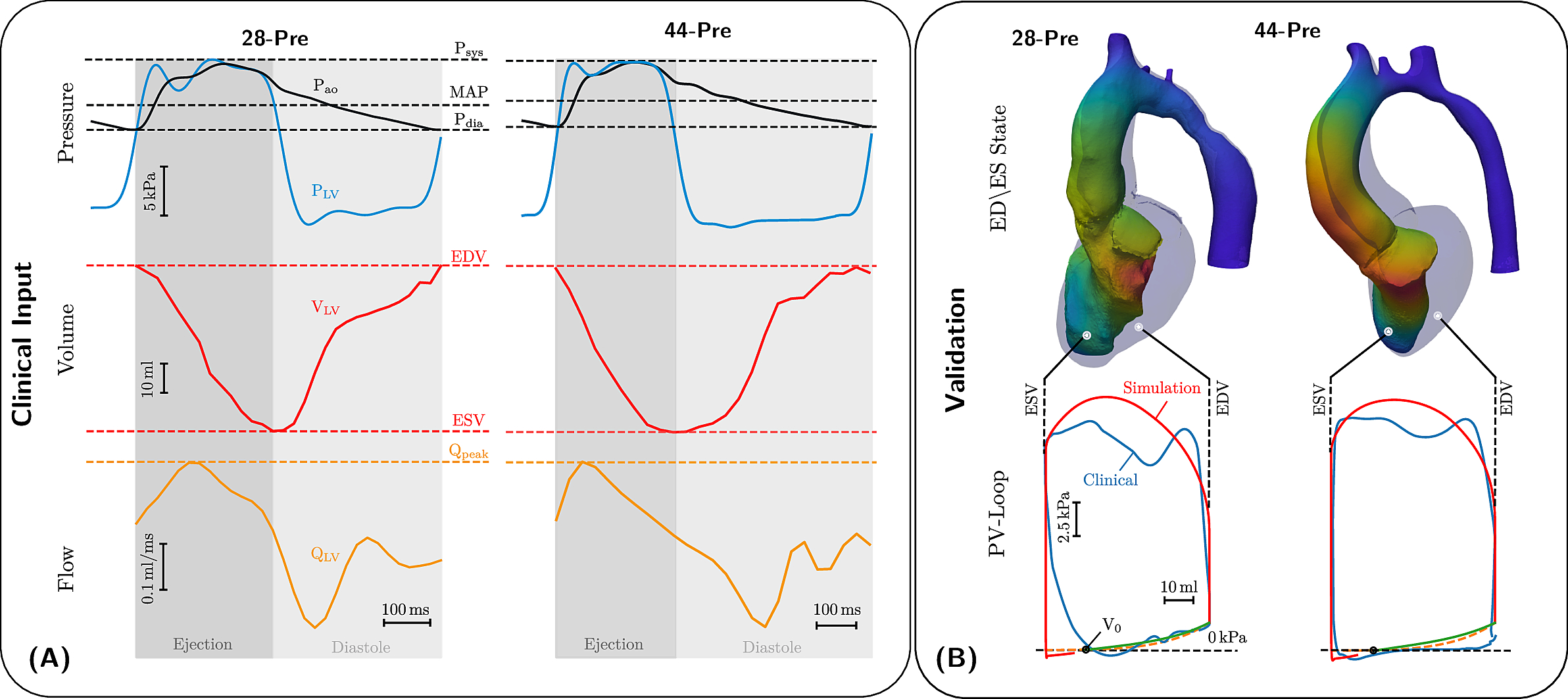}
  \caption{\textsf{(A)} Invasive clinical recordings from cases \emph{28-Pre} and \emph{44-Pre}.
    Top: Recorded aortic pressure $\mathrm{P}_\mathrm{ao}$ (black curve)
      and recorded LV pressure  $\mathrm P_\mathrm{LV}$ (blue curve).
      Marked with dashed lines are Systolic pressure $\mathrm{P}_\mathrm{sys}$,
      mean arterial pressure $\mathrm{MAP}$, and
      diastolic pressure $\mathrm{P}_\mathrm{dia}$;
      Center: Volume change in the LV, $\mathrm V_\mathrm{LV}$, in red ranging
      from end-diastolic volume $\mathrm{EDV}$ to end-systolic volume $\mathrm{ESV}$.
    Bottom: LV flow $\mathrm{Q}_\mathrm{LV}$ in orange with marked peak flow $\mathrm{Q}_\mathrm{peak}$.
     \textsf{(B)} Comparison of EM simulations and clinical data.
      Upper part shows a comparison of the LV model in end-diastolic (colored opaquley blue)
      and end-systolic configuration (colored by displacement).
      Lower part shows comparison of clinical (colored blue) and simulated PV loops (colored red).
      The dashed orange curve shows the ideal Klotz curve,
      while the green curve shows the simulated Klotz curve, with volume of stress-free unloaded configuration marked as $\mathrm V_0$.}
  \label{fig:_results_EM_validation}
\end{figure}

Passive biomechanical properties, afterload and active stress models of cases
\emph{28-Pre} and \emph{44-Pre} were parameterized using clinically recorded
pressure and volume data under pre-treatment conditions, see \autoref{fig:_results_EM_validation}A.
The fitted final parameters used are summarized in \autoref{tab:_EM_param_fit}.
The goodness of fit of both integrated EM models was verified by standard PV
loop analysis as shown in \autoref{fig:_results_EM_validation}B.
Results of a quantitative comparison with clinically derived metrics including
EF, EDV and ESV, CO, and peak systolic pressure are summarized in \autoref{tab:_EM_validation}.
%
%------------------------------------------------------------------------------
\subsection{Blood pool FE modeling for CFD}
\label{sec:res:bloodpool:CFD}
Conformal FE blood pool meshes were extracted from EM FE meshes, surfaces were smoothed
and used for volumetric remeshing with increased spatial resolution including boundary layers.
The corresponding workflow is illustrated in \autoref{fig:fluid_model_generation}.

\begin{figure}[htb]
  \centering
  \includegraphics[width=\textwidth,keepaspectratio]{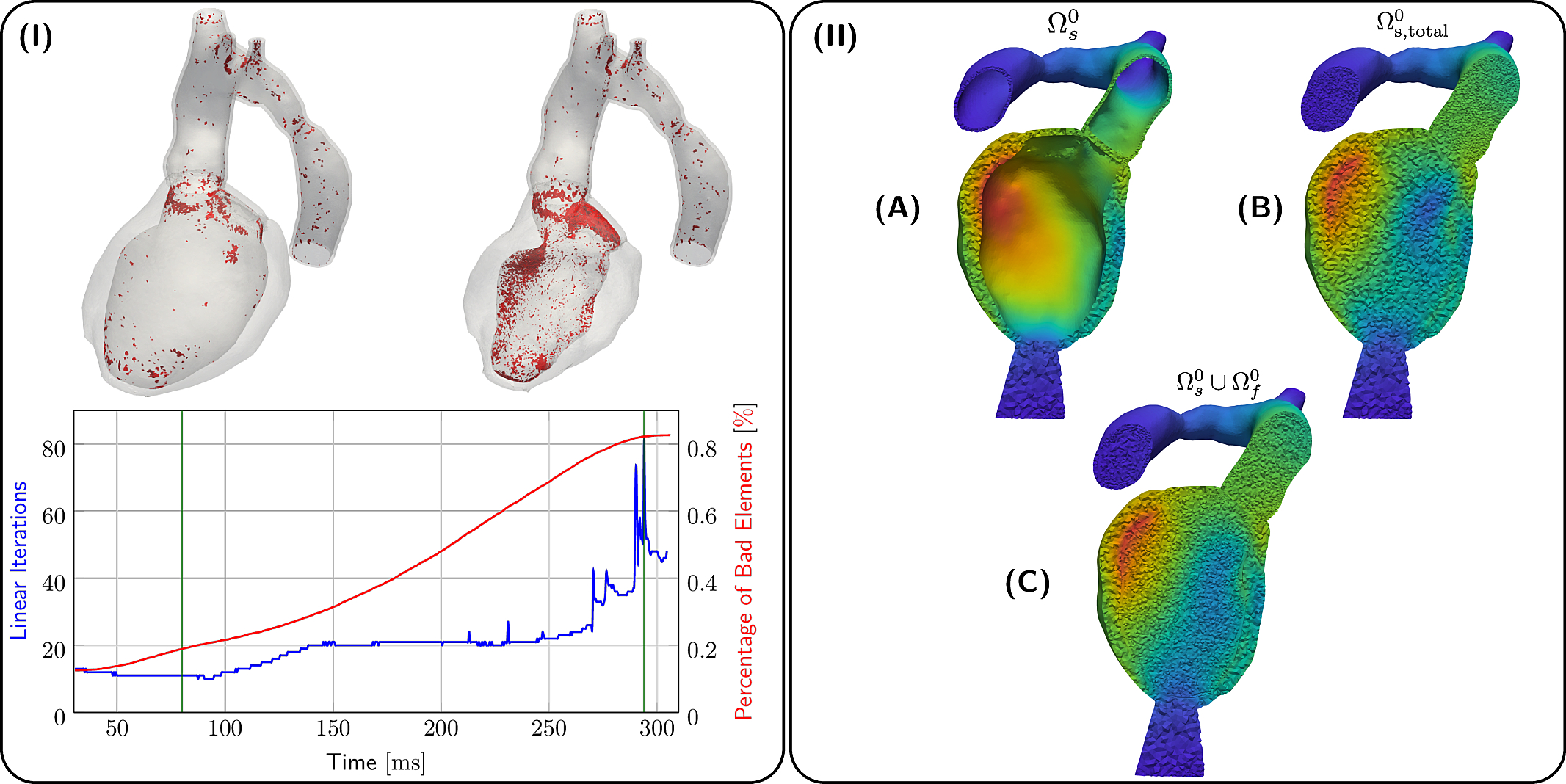}
  \caption{Panel \textsf{(I)} shows quality analysis for case \emph{28-Pre}.
   Spatial locations of elements of poor quality $>0.8$ (in red) are shown at the
   top for different snapshots of deformation (green lines in graph).
   The graph below shows linear iterations per time step (in blue) and percentage of
   elements with poor quality $>0.8$ (in red).
   Panel \textsf{(II)} shows the processing stages of kinematic transfer for the
   \emph{28-Pre} case at maximum displacement.
   \textsf{(A)} Displacement $\vec d_\mathrm{s}$ on EM mesh $\Omega_\mathrm{s}^0$.
   \textsf{(B)} Displacement $\vec d_\mathrm{s}$ extended to conformal EM blood pool
   mesh $\Omega_\mathrm{s,total}^0$ which serves as hanging background mesh
   for the kinematic transfer onto the CFD blood pool mesh $\Omega_\mathrm{f}^0$.
   \textsf{(C)} Displacement $\vec{d}_\mathrm{s}$ on $\Omega_\mathrm{s}^0$
   superimposed with fluid mesh displacement $\vec{d}_\mathrm{f}$ on
   $\Omega_\mathrm{f}^0$.
  }
  \label{fig:quality_analysis}
\end{figure}

Kinematics of the EM model were transferred to the CFD blood pool mesh and the
result is illustrated in terms of displacements
$\vec d_\mathrm{s}, \vec d_\mathrm{f}$ in Panel \textsf{(II)} of
\autoref{fig:quality_analysis}.
Due to the large EF of about $\SI{65}{\percent}$ for both \emph{28-Pre} and \emph{44-Pre},
the blood pool underwent a significant deformation.
However, using a combination of element quality and $\nu$-Volume based stiffening
with an initial Young's Modulus $E_0 = \SI{100}{\kPa}$ and Poisson's ratio $\nu_0 = 0.3$,
sufficient element quality was preserved throughout the entire ejection phase %\todoCA{throughout the cardiac cycle?}
and numerical instabilities could be avoided.
Panel \textsf{(I)} of \autoref{fig:quality_analysis} shows the
$80^\mathrm{th}$-percentile of bad element quality against the number of linear
iterations required for convergence for the \emph{28-Pre} case. The quality of elements was calculated with the same
quality inidcator \cite{freitag2002tetrahedral,kanchi20073d} as described in Section \ref{sec:transfer} but was rescaled to the interval
$[0,1]$, with the best element quality being $0$ and the worst element quality being $1$.
The modest increase in iteration numbers of the iterative preconditioned GMRES solver
provides indirect evidence
of sufficiently preserved mesh quality (see \autoref{fig:quality_analysis}).
Spatially, most lower quality elements were located in the CFD boundary layer.
%
%------------------------------------------------------------------------------
\subsection{Numerical CFD benchmarks}
\begin{figure}[htb]
\centering
\includegraphics[height=8cm,keepaspectratio]{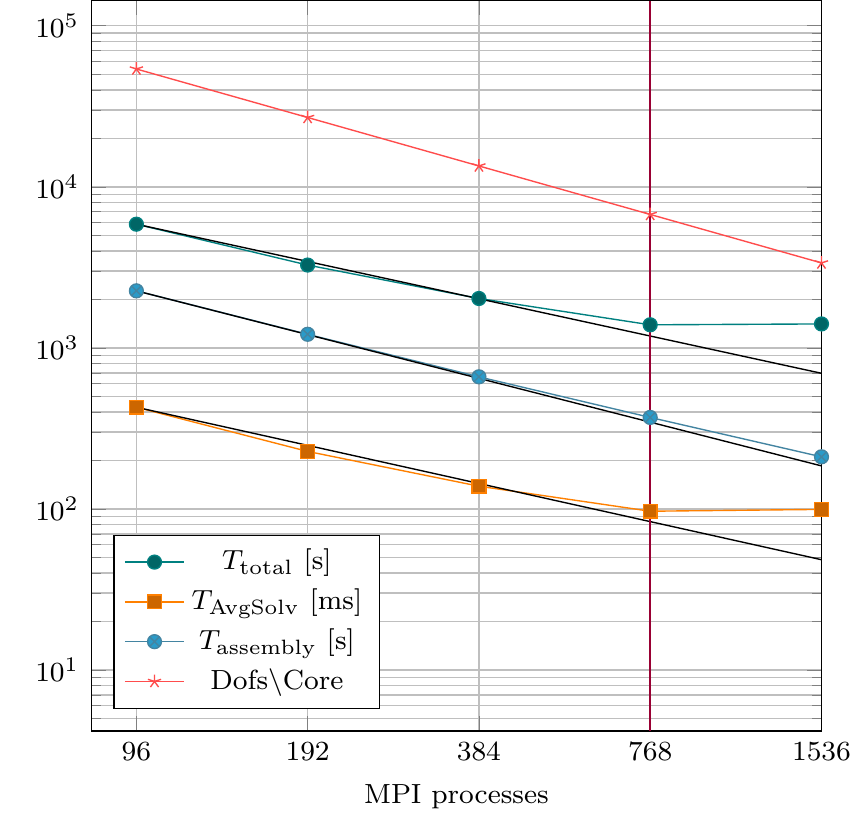}
\caption{Results of strong scaling benchmark based on case \emph{28-Post}
with \num{5.2} million overall degrees of freedom.
$T_\mathrm{AvgSolv}$ is the total solving time
divided by the total amount of linear iterations per simulation run.}
\label{fig:scaling}
\end{figure}

The implementation of the Navier--Stokes solver was verified
by solving a set of standardized benchmark problems, see \cite{schafer1996benchmark}.
Computational performance was evaluated by performing strong scaling experiments
by repeating the post-treatment hemodynamics simulation of case \emph{28-Post}
with varying numbers of cores ranging from \num{96} to \num{1536}.
Details on computational complexity and costs are summarized in \autoref{tab:meshes}.
For temporal discretization a time step of $\Delta t=\SI{0.5}{\milli\second}$ was used
to simulate the ejection phase lasting for \SI{208}{\milli\second}.
The overall discrete system comprised \num{5177056} degrees of freedom,
which was solved over \num{416} time steps.
Strong scaling results are summarized in \autoref{fig:scaling}.
Efficient strong scaling behavior was observed up to \num{768} cores
with parallel efficiency slowly degrading from \SI{100}{\percent} at \num{96}
cores down to \SI{55}{\percent} at \num{768} cores.
Scalability stalled when doubling the core count to \num{1536}
which reduced the degrees of freedom per parallel partition down to \num{3386}.
Parallel efficiency dropped to \SI{27}{\percent}
which is attributed due to the unfavorable ratio between local compute work and communication.
%
%------------------------------------------------------------------------------
\subsection{Simulating cardiac and cardiovascular hemodynamics}\label{sec:hemo1}
\begin{figure}[htb]
  \centering
  \includegraphics[width=\textwidth,keepaspectratio]{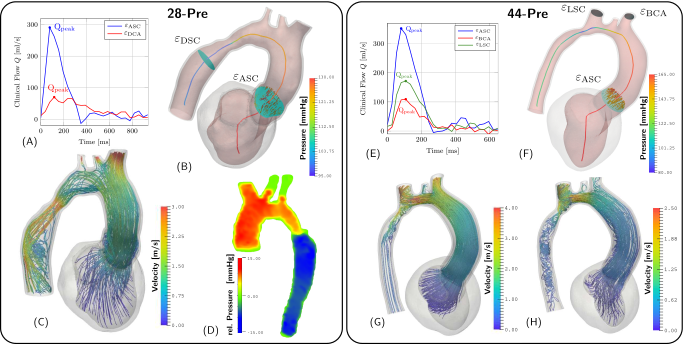}
  \caption{CFD results. Subfigures \textsf{(A)}, \textsf{(E)} show the given clinical measurements for flow through different planes. The planes are depicted in Subfigures \textsf{(B)} and \textsf{(F)}. Subfigures \textsf{(B)} and \textsf{(F)} also depict the pressure along the centerlines at peak flow conditions at $t=\SI{167}{\ms}$ and $t=\SI{142}{\ms}$ respectively. Subfigure \textsf{(C)} shows velocity streamlines at peak flow. Subfigure \textsf{(D)} shows the relative pressure map from the Pressure--Poisson mapping used for validating the pressure drop in our simulations. Subfigures \textsf{(G)}, \textsf{(H)} show velocity streamlines at peak flow and $t=\SI{200}{\ms}$ for case \emph{44-Pre}.}
  \label{fig:cfd_validation}
\end{figure}

Hemodynamics in the LV and aorta was simulated using the EM simulations as a kinematic driver.
Flow rates through various aortic cross sections and outflow orifices were calculated
as the integral over measured fluxes through the cross-sectional plane for both 4D VEC MRI and simulated flow data.
At locations of interest which were
$\varepsilon_\mathrm{DSC}$, $\varepsilon_\mathrm{BCA}$, $\varepsilon_\mathrm{LCA}$ and $\varepsilon_\mathrm{LSCA}$
denoting cross sections in the aorta descendens and the orifices of brachocephalic, left carotid and left subclavian artery, respectively,
relative flows were computed from 4D VEC MRI data
as fractions $\alpha_i$ expressed in percent of the total peak flow
through the aorta ascendens as determined over the plane  $\varepsilon_\mathrm{ASC}$.
For those planes of interest where measurements were not feasible due to noise,
flow percentages  were estimated based on Murray's law.
Flow curves during ejection at selected cross sections are shown in Subfigures \textsf{(A)}, and \textsf{(E)} of \autoref{fig:cfd_validation}.
MAP and computed mean flow through each outlet orifice were used
to determine the parameters of the coupled Windkessel models of afterload in
Equations~(\ref{eq:cfd_wk:1},~\ref{eq:cfd_wk:2}),
see Tables~\ref{tab:WK28} and \ref{tab:WK44}.
In the \emph{28-Pre} case this resulted in flow splits of
$\alpha_i\approx \SI{23}{\percent}, \SI{51.3}{\percent}, \SI{12.83}{\percent}$ and $\SI{12.83}{\percent}$
whereas in the \emph{44-Pre} case the flow split ratios were
$\alpha_i\approx  \SI{5.68}{\percent}$, $\SI{57.45}{\percent}$ and   $\SI{34.01}{\percent}$
for $\varepsilon_\mathrm{DSC}$, $\varepsilon_\mathrm{BCA}$, $\varepsilon_\mathrm{LCA}$
and $\varepsilon_\mathrm{LSCA}$, respectively.

For the CFD analysis a time step of $\Delta t = \SI{0.5}{\milli\second}$ was used.
The ejection phases of the EM simulations were chosen as time horizons for the CFD simulation
which lasted from $t=\SI{90}{\milli\second}$ to $t=\SI{302}{\milli\second}$ in the \emph{28-Pre} case
and from $t=\SI{70}{\milli\second}$ to $t=\SI{329}{\milli\second}$ in the \emph{44-Pre} case,
yielding 424 and 518 time steps, respectively.
The Windkessel parameters for each outlet, calculated as
described in Section~\ref{sec:cfd_boundary_conditions},
are summarized in Tables~\ref{tab:WK28} and \ref{tab:WK44}.
Pressure $p_\mathrm{f}$ along the centerline $s_{\rm c}$ and fluxes through the planes $\varepsilon_\mathrm{DSC}$, $\varepsilon_\mathrm{LSC}$, $\varepsilon_\mathrm{BCA}$, and $\varepsilon_\mathrm{ASC}$ were computed at the instant of peak flow in the aorta ascendens and compared against measured data,
which were pressures derived from Pressure--Poisson mapping (see Subfigure \textsf{(D)} of \autoref{fig:cfd_validation}) and 4D VEC MRI fluxes.
For case \emph{28-Pre} pressure drops were calculated from the pressure values on the intersection of the centerline and $\varepsilon_\mathrm{DSC}$, $\varepsilon_\mathrm{ASC}$ respectively.
Further, we calculated the average pressure over the aforementioned planes as well. Both ways yielded a simulated pressure drop across the CoA of $\approx \SI{29.2}{\mmHg}$ which agreed well with the clinically estimated pressure drop of $\approx \SI{30}{\mmHg}$.
Furthermore, we calculated the flux through the various planes and compared them against the clinically estimated fluxes.
A quantitative comparison of fluxes is given in \autoref{tab:flux_comp}.
Subfigures \textsf{(C)}, \textsf{(G)}, and \textsf{(H)} of \autoref{fig:cfd_validation} show velocity profiles at peak flow condtions.

\subsection{post-treatment simulations}
\begin{figure}[htb]
  \centering
  \includegraphics[width=\textwidth,keepaspectratio]{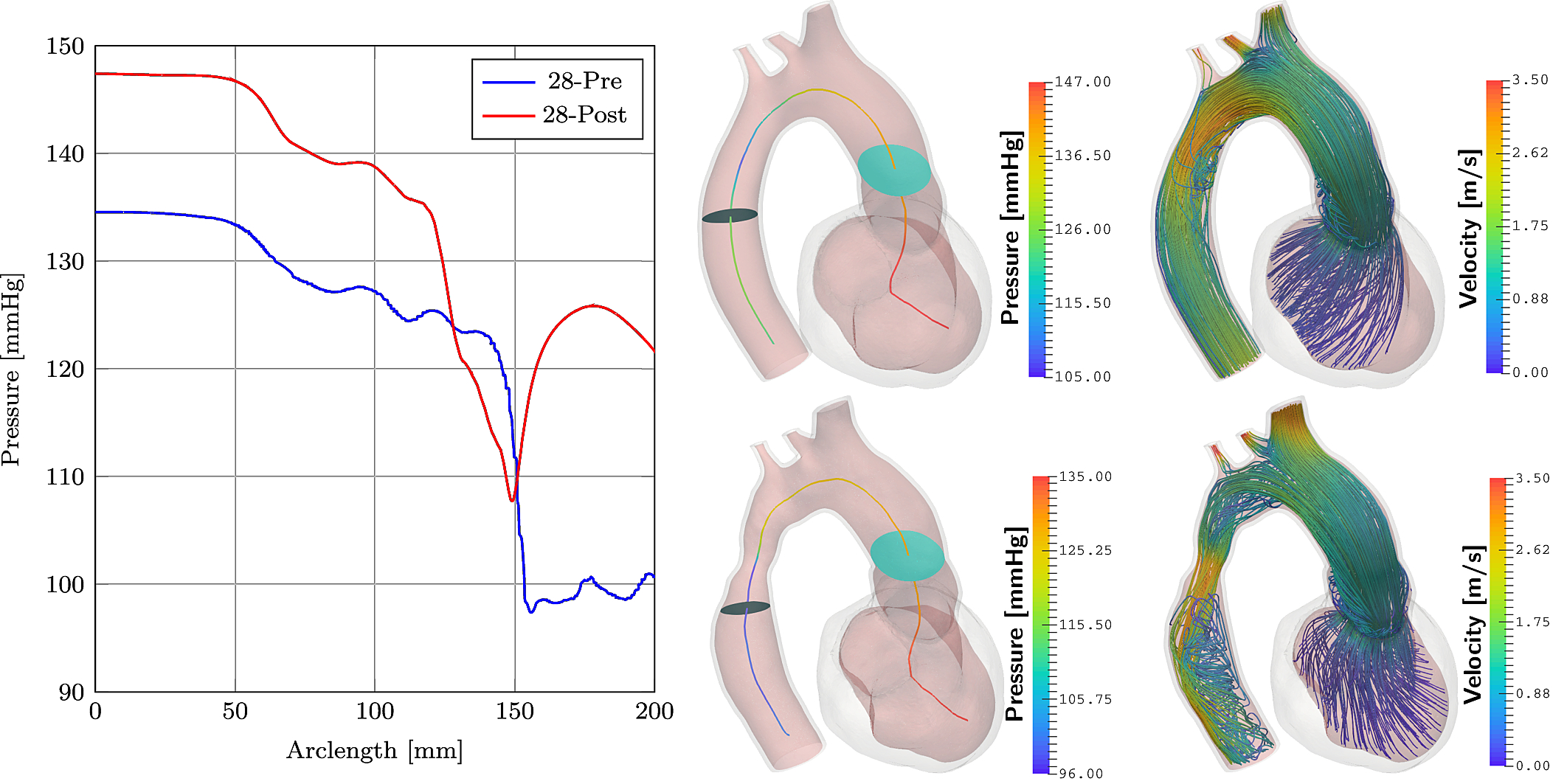}
  \caption{Comparison of cases \emph{28-Pre} and \emph{28-Post}. Shown on the left are the pressures along the centerline at peak flow. Depicted in the middle are the slices used for calculating the pressure drops. Shown on the right are velocity streamlines at peak flow.}
  \label{fig:post_treatment_results}
\end{figure}

Simulations of case \emph{28-Pre} were repeated on geometry of case \emph{28-Post} using almost the same set of parameters, see \autoref{tab:_EM_param_fit}. Only $S_\mathrm{peak}$ was slightly adjusted, which resulted in a better peak pressure value in the LV. The geometry of case \emph{28-Post} was almost identical to case \emph{28-Pre} with the only exception being the virtual repair of CoA anatomy.
In this scenario only pre- and post-treatment simulations were compared
to evaluate their relative differences in terms of pressure and flow velocities. \autoref{fig:post_treatment_results} shows results. Pressure drops were calculated as in Subsection \ref{sec:hemo1} for both scenarios. For \emph{28-Pre} we calculated a pressure drop of $\approx \SI{29.2}{\mmHg}$ while for \emph{28-Post} a pressure drop of $\approx \SI{14.15}{\mmHg}$ was calculated.

\section{Discussion}

In this study, we report on the progress made towards a novel EMF model of the human LV
that is entirely based on first principles and as such,
in principle, is able to represent all cause-effect relationships with full biophysical detail.
Unlike in the majority of cardiac CFD studies
where the use of image-based kinematic driver models prevails,
EM LV and aorta models of CoA patients were employed
to serve as a kinematic driver to a computational model of hemodynamics in the LV cavity and aorta.
% and is able to represent all cause-effect relationships with full biophysical detail.
A hybrid two stage modeling approach was adopted with regard to hemodynamics
where EM and CFD model are executed sequentially.
First, in the EM simulations the afterload imposed by the circulatory system upon the LV
was represented by a lumped model to compute LV kinematics.
These EM models were carefully fitted to available clinical data
to replicate important clinical metrics characterizing hemodynamic and biomechanical work performed by the LV \cite{gsell17:_assessment}.
In a subsequent step, a full-blown ALE-based CFD model with moving domain boundaries
was \emph{unidirectionally} or \emph{weakly} coupled to the EM model.
The motion of the fluid domain was driven by the kinematics of the EM model.
Kinematics was transferred from EM mesh onto the CFD blood pool mesh
by generating a combined kinematic model comprising LV, valve, aortic structure
and a conformal blood pool mesh which served as a hanging background mesh for interpolation.
The higher resolution blood pool CFD mesh with refined boundary layers
was fully immersed in the EM background mesh.
Kinematics was transferred by interpolation only onto the surface of the CFD blood pool mesh
and extended into the volume of the blood pool by solving a linear solid mechanics problem.

We show validation results for two selected clinical CoA cases under pre-treatment conditions
and compare between pre-treatment and post-treatment for one patient case
in which the CoA was anatomically modified by a virtual stenting procedure.
Further, we demonstrate numerical tractability of the implemented approach
by providing strong scaling benchmark results.
The overall cost of the entire work flow for building, fitting and execution of EMF simulations
is comparable to plain image-based kinematic driver models \cite{mittal16:_computational},
suggesting that the proposed methodology may be, in principle, compatible with clinical time scales.
%
%------------------------------------------------------------------------------
\subsection{Biomechanical modeling versus image-based kinematics}
Modalities such as CMR and Cardiac CT on the other hand, provide excellent
spatial resolution. CMR has an in-plane resolution of
1.5 $\times $ \SI{1.5}{\mm}, but more limited through-plane resolution
(typically about \SI{8}{\mm}) while CT is capable of isotropic spatial resolution
on the sub millimeter scale ($\approx \SI{0.5}{\mm}$)
and clear delineation of trabeculae and lumen boundaries.
CMR has the advantage of higher temporal resolution (\SIrange{30}{50}{\ms})
while temporal resolution in CT depends on the scanning system
(\SIrange{50}{200}{\ms}).
This is orders-of-magnitude lower than the temporal resolution required for the flow
simulation ($\approx 1000$ phases per cardiac cycle) and appropriate
interpolation methods need to be employed to create CFD-ready models.
This stage of model generation has been very difficult to automate,
and remains the biggest bottleneck for patient-specific cardiac flow modeling.
Compared to pure image-based kinematic approaches our model is able to compute, e.g.,
the spatio-temporal distribution of wall stresses, power density,
the length of diastolic intervals available for myocardial perfusion,
$\mathrm{O}_2$ consumption, and metabolic supply/demand ratios.
The variations of all these parameters in response to a changed afterload
and many other biomarkers of physiological interest can be derived,
which is not feasible with image-based models.
%
%------------------------------------------------------------------------------
\subsection{Kinematic transfer to CFD blood pool model}
Both patients modeled in this study featured healthy EFs of
$> \SI{60}{\percent}$, that is, EF was $\approx \SI{65}{\percent}$ in both cases.
At a such high EFs the wall motion of the LV is significant,
leading to substantial reductions in the LV blood pool volume.
IB methods \cite{choi2015anew,seo2013effect,vigmond2008effect}
are known to be more convenient to cope with the large deformation of the CFD
blood pool \cite{quarteroni2017integrated}.
IB methods and other non-boundary-fitting methods rely on a fixed fluid mesh
and the moving wall of the ventricle is not explicitly tracked.
The coupling between the CFD mesh and the structure is performed via Dirac Delta
functions (IB) or Lagrange multipliers (fictitious domain methods) and is
usually realized by introducing additional degrees of freedom on interface cut elements.
While mesh generation is only necessary prior to computation fixed mesh
methods typically require adaptive mesh refinement or modifications
\cite{wang2004extended} to obtain reasonable accuracy for the solution near
the fluid-solid interface.

In contrast, ALE algorithms capture the fluid-solid interface more accurately,
are in general stable and easy to implement, no extra degrees of freedoms are introduced,
and computational costs are low in comparison
\cite{vanloon2007comparison,letallec2001fluid}.
However, it is often assumed that unstructured FE approaches, as implemented in this paper,
critically depend on automatic remeshing strategies \cite{long2013fluid}
to keep mesh quality within acceptable bounds \cite{mittal16:_computational}.
Our study demonstrates that this may not necessarily be the case.
While the mesh quality decreased with deformation over the course of ejection,
the linear elastic deformation of the CFD blood pool mesh combined with the quality-based stiffening approach
prevented the degeneration of any elements.
The number of elements in which element quality degraded noticeably was very small.
As illustrated in \autoref{fig:quality_analysis}, virtually all elements of reduced quality
were located in the higher resolution boundary layer of the CFD blood pool mesh.
According to the element quality metric used, an element quality of 1
refers to a fully degenerated element of zero volume.
Despite the significant compression of the blood pool mesh,
not a single element was deformed to this degree.
Even when applying a stricter threshold
where element quality is deemed poor if the quality indicator is $>0.8$,
which is not critical from a numerical point of view,
the number of elements in this range remained small with $<\SI{0.8}{\percent}$
(\autoref{fig:quality_analysis}).
The worst element quality observed in the entire mesh was $0.9994$.
Using a threshold of $>0.95$ where element quality may be sufficiently poor
to impact more notably on solver performance, only 24 out of 2506987 elements were found.
Nonetheless, an increase in number of linear iterations required for convergence
was observed
which is likely to be linked to the gradual degradation of element quality.
The number of iterations per solver step increased from around
$\approx 17$ iterations during early ejection
up to $\approx 80$ iterations during late ejection.
While the more than fourfold increase in linear iterations negatively impacted overall solver performance
and rendered simulations computationally more expensive,
the complexity of automatic remeshing was avoided.
We consider this a pivotal importance as automatic remeshing in combination
with a MPI parallel FE solver is definitely feasible,
but highly non-trivial to implement robustly and efficiently.

% Will stay open until revision.
% \todoCA{Do we want a comparison to these papers as well? They are
%    ALE only but rely on simplifications and I mentioned them in the
%    introduction
%    \begin{itemize}
%        \item \citet{vazquez2015alya} weakly and fully coupled (the latter only
%          preliminary); no validation
%        \item \citet{Tang2010,tang2008patient} rather RV
%        \item \citet{devecchi2016novel,nordsletten2011fluid} ALE + IB for valves
%    \end{itemize}
%    Papers not yet cited:
%    \begin{itemize}
%        \item \citet{Nordsletten2011} review
%        \item \citet{lassila2017simulation} fictitious elastic structure
%        \item \citet{basting2017extended} Extended ALE $\rightarrow$ large deformations without
%            changing mesh connectivity; price to pay: mesh optimization involves
%            solving a highly non-liear, non-convex, and non-unique problem;
%            probably we could use similar techniques to improve the mesh and
%            model valves;
%        \item surprisingly I didn't find any papers from Gerbeau, Chapelle, Moireau and INRIA-Konsorten
%              aber man kann nicht alles zitieren :)
%    \end{itemize}
%}

%Nonetheless, it remains to be tested whether remeshing pays off in terms of reduced execution times.
%While it is likely feasible to keep iteration numbers low over the entire ejection,
%remeshing in a FEM context
%
%------------------------------------------------------------------------------
\subsection{Computational feasibility}
Computational feasibility of human scale cardiac simulations
by using strongly scalable numerical implementations has been demonstrated previously
for electrophysiology \cite{niederer11:_clinical} and mechanics \cite{augustin16:_anatomically}.
More recently, we reported on a novel reaction-eikonal model
which reduces the cost of EM simulations significantly
by alleviating constraints imposed by reaction-diffusion models upon mesh resolution \cite{neic17:_efficient}.
In this study, this recent reaction-eikonal approach was used for simulating EM
using the same FE grid with an average resolution of $\approx 1$ mm for both EP and mechanics.
Such lower resolutions suffice for solving for mechanics with sufficient accuracy \cite{land15:_nversion}.
The overall reduction in terms of nodes and degrees of freedom reduces the compute cost substantially,
rendering simulations in desktop environments feasible.
Using 96 cores, EM simulations of a full cardiac cycle only lasted $\approx \SI{180}{\minute}$
which facilitated sufficiently short simulation cycles for efficient model fitting.
The entire workflow for building and parameterizing one patient-specific EM model is feasible within a day.

Owing to the higher resolution of the blood pool mesh and the presences of a refined boundary layer
the number of nodes and degrees of freedom were higher than for EM simulations,
around $350000/1500000$ nodes/degrees of freedom for case \emph{28-Pre}
and $400000/1700000$ nodes/degrees of freedom for case \emph{44-Pre}, respectively.
To assess strong scaling properties of our CFD solver implementation,
the resolution was further increased to $1300000/5000000$ nodes/degrees of freedom for case \emph{28-Post}
to cover a wider range of core counts.
Strong scaling efficiency leveled off when doubling from 768 to 1536 cores.
Local compute load with 1536 was $900/2600$ nodes/degrees of freedom per core.
The patient simulations were performed using 384 cores,
resulting in a load per core of about $900/2700$ nodes/dofs, respectively.
At these resolutions CFD simulations were executed in $\approx \SI{40}{\minute}$,
suggesting that compatibility with clinical time frames will be achievable.
%
%------------------------------------------------------------------------------
\subsection{Limitations}
In the presented modeling approach numerous simplifying assumptions were made
which may affect the biophysical fidelity of the model.
In particular, while the aorta was taken into account as a solid structure in the EM simulations,
its biomechanical description was simplified by assuming isotropic behavior,
that is, the fibrous organization of  aortic walls remained unaccounted for
\cite{augustin2014classical}.
Further, as our main focus was on the EM of the LV and, to a much lesser degree, on the aorta,
the aortic lumen remained unpressurized and, in absence of distensibility measurements of the aortic wall,
parameters of the passive biomechanics model used for the aortic wall were not fitted.
Thus the model of the aorta does not respond to the rise in pressure during ejection
with an adequate distension $\Delta V$ of its lumen.
In the CFD simulations $\Delta V \approx 0$ translates into a stiff aorta of low compliance
which may cause a bias towards overestimation of the computed pressure fields.
Further, the influence of the aortic valve upon blood flow was not taken into account.
Rather, it was assumed that with the start of ejection the aortic valve is in its full open configuration,
which allows blood flow over the entire orifice area
and in which the valve does not influence the blood flow out of the LV in a significant way.
Since only CoA patients were modeled which showed no indications of AVD
this simplifying assumption may be well justified.

A potential main strength of the presented modeling approach
-- the ability to predict the biomechanical response of the LV to changed flow patterns in the aorta --
was not exploited.
Due to the weak FSI coupling the immediate feedback of altered flow
or changed pressure gradients in the aorta on LV biomechanics was ignored.
In our current modeling approach any such feedback must be mediated through changes
in the parameterization of the lumped afterload model.
However, owing to regulatory mechanism of the circulatory system level
this is not directly predictable with the modeling setup used in this study
as flow distribution through the four outlets will be influenced by factors
which cannot be accounted for in a model comprising only LV, aorta and lumped outflow impedances.
In any case, one cannot assume that the computed changes in pressure gradients across a CoA
translate directly into a reduction in LV peak pressure.
Independently of the modeling approach taken
-- be it a strongly or weakly coupled FSI model --
a lumped model of systemic regulation is likely to be necessary to predict altered LV loading
under post-treatment conditions \cite{arts2005adaptation,lumens2009threewall}.
Compared to a fully coupled FSI model our approach is limited in the sense
that CFD simulations do not influence the behavior of the EM model.
However, in many clinical settings CFD simulations
in the aortic arch and LV with image based kinematics prevail.
Image based kinematic models can only depict the \emph{status quo} of a patient.
With our personalized EM model, based on first principles,
we can do simulations altering the motion, simply by changing input paramters.
The altered motion is then reflected in the CFD simulation. Examples would include changes in heart beats, infarcts or LBBB conditions.
In this work, the effect of stenting was only accounted for by a geometric change in the
computational geometry and an ad hoc adjustement of the lumped model parameters.
In future studies, we intend to use a 1-D model of the arterial tree coupled to a
0-D lumped model at the aortic outlets, thus being able to account for the effect
of stenting in a more detailed fashion, see for example \cite{quarteroni2017integrated}.
As a first step towards our ultimate goal of a fully coupled FSI model, that is based entirely on first principles, we will add the dynamic fluid pressure $\frac{\rho_\mathrm{f}}{2}\abs{\vec u_\mathrm{f}}^2$ to the pressure of the lumped model (0-D or 1-D).
This results in a spatio-temporal pressure inside the LV and the aorta, and to
incorporate the dynamic feedback of fluid upon structure we will iterate between
a CFD solving step and a EM solving step within each timestep to guarantee a
converged solution.

%
%------------------------------------------------------------------------------
\section{Conclusion}
Biophysically detailed models of LV EM can be efficiently built and parameterized with clinical data
to be considered a viable option for patient-specific simulation.
Similar to image-based kinematic models such biophysics-based EM models can be used as a kinematic driver
for simulating cardiac and vascular hemodynamics.
The cost of model building and execution is comparable between the two approaches.
Biophysical EM models offer the significant advantage of being based entirely on first principles
and as such, may allow to make predictions of interventions altering pressure and flow patterns onto LV performance.
In contrast, image-based kinematics modeling may provide a more accurate representation of blood pool motion,
at least under pre-treatment conditions or post-treatment conditions secondary to interventions
which do not influence LV kinematics in a significant way.
%
%------------------------------------------------------------------------------
\section*{Conflict of Interest Statement}
%All financial, commercial or other relationships that might be perceived by the academic community as representing a potential conflict of interest must be disclosed. If no such relationship exists, authors will be asked to confirm the following statement:
The authors declare that the research was conducted in the absence of any
commercial or financial relationships that could be construed as a potential conflict of interest.
%
%------------------------------------------------------------------------------
\section*{Author Contributions}
EK, GP contributed conception and design of the study;
LG, TK acquired and processed clinical data;
EK, MG, AN and CA developed numerical methodology;
AJP contributed by conceiving modeling workflows and FE meshing;
LM and MG developed parameterization of electromechanical model;
EK, MG, CA and GP analyzed and interpreted simulation data;
EK, CA and GP  drafted the article;
EK, CA, MG, LG and GP critically revised the article;
All authors contributed to manuscript revision, read, and approved the submitted version.
%
%------------------------------------------------------------------------------
\section*{Funding}
This research was supported by the grants F3210-N18 and I2760-B30
from the Austrian Science Fund (FWF), the EU grant CardioProof agreement 611232
and a BioTechMed award to GP,
and a Marie Sk{\l}odowska--Curie fellowship (GA 750835) to CA.
We acknowledge PRACE for awarding us access to resource ARCHER based in the UK at EPCC
(grant CAMEL) and the Vienna Scientific Cluster VSC-3.

%\printbibliography
\bibliographystyle{plainnat}
\bibliography{references}

\clearpage
\section*{Tables}

%% tab1
\begin{table}[htb]
  \ra{1.3}
  \centering
  \begin{tabular}{@{}lccccccccccccc@{}}
    \toprule
    & Sex & Age   & EDV                  & ESV & SV
    & EF & HR & CO & $P_\mathrm{ao/cuff}^{\mathrm{dia}}$
    & $P_{\mathrm{ao/cuff}}^{\mathrm{sys}}$
    & MAP & $P_{\mathrm{open}}$\\
    & &  & \si{\ml}  & \si{\ml}   & \si{\ml}    & \si{\percent}  & $\mathrm{bpm}$
    & \si{\ml\per\s} & \si{\mmHg} & \si{\mmHg}  & \si{\mmHg} & \si{\mmHg} \\
    \midrule
    \emph{28-Pre}  & F &  9 &  88.2 & 30.6 &  57.54 & 65.3 & 91 & 87.46
                   & $\sfrac{71.1}{62}$    & $\sfrac{122.7}{138}$
                   & $\sfrac{88.3}{87.3}$   & $71.33$ \\
%    \emph{28-Post} & F &  9 &  65.0 & 32.1 &  32.93 & 50.6 & 59
%                   & $\sfrac{70.7}{51}$    & $\sfrac{109.8}{100}$
%                   & $\sfrac{83.7}{67.3}$   \\
    \emph{44-Pre}  & M & 12 & 91.7 & 31.6 & 60.09 & 65.5 & 76 & 76.31
                   & $\sfrac{74.6}{120}$ & $\sfrac{125.2}{154}$
                   & $\sfrac{91.5}{131.3}$ & $74.78$ \\
    \bottomrule
  \end{tabular}
  \caption{CoA patient characteristics from MRI and invasive catheter pressure
    recordings including end-diastolic volume (EDV), end-systolic volume (ESV),
    stroke volume (SV), ejection fraction (EF), heart rate (HR),
    cardiac output (CO), diastolic and systolic pressures recorded in the
    aorta or estimated from cuff measurements ($P_{\mathrm{ao/cuff,dia}}$
    and $P_{\mathrm{ao/cuff,sys}}$),
    mean arterial pressure (MAP) computed from pressure recorded invasively
    in the aorta or estimated from $P_{\mathrm{cuff,dia}}$
    and $P_{\mathrm{cuff,sys}}$,
    and aortic valve  open pressure $P_{\mathrm{open}}$ determined from
    invasive pressure recordings.
  }
  \label{tab:patients}
\end{table}

%% tab2
\begin{table}[htb]\centering
\ra{1.3}
\begin{tabular}{@{}cr@{}}\toprule
\textbf{Tag} & \textbf{Label} \\
\hline
$\mathrm{lv}$  & Myocardium \\
$\mathrm{ao}$  & Aortic wall \\
$\mathrm{cushion}$ & Elastic cushion \\
$\mathrm{av}$ & Aortic valve \\
$\mathrm{mv}$ & Mitral valve \\
$\mathrm{lvbp}$ & Left ventricular bloodpool \\
$\mathrm{aobp}$ & Aortic bloodpool \\
\bottomrule
\end{tabular}
\caption{Labels used for defining the subdomains of $\Omega_{s,\mathrm{total}}^0$.}
\label{tab:taglabels}
\end{table}

%% tab3
\begin{table}[htb]
	\ra{1.3}
	\centering
	\begin{tabular}{@{}lccccccccc@{}}
		\toprule
		& \multicolumn{9}{c}{\textbf{EM Fitting}}  \\
		\hline
		& $S_\mathrm{peak}$ & $t_\mathrm{dur}$& $\tau_{\mathrm{c}0}$ & $\tau_\mathrm{r}$ & $t_\mathrm{emd}$& $C_\mathrm{Guc}$ & $R$ & $Z$  & $C$ \\
		& $\si{\kPa}$ & $\si{\ms}$ & $-$ & $-$ & $\si{\ms}$ & $\si{\kPa}$ & $\si{\kilo\Pa \milli\second\per\milli\litre}$  &  $\si[per-mode=symbol]{\kilo\Pa \milli\second\per\milli\litre}$  & $\si{\milli\litre\per\kilo\Pa}$ \\
		\midrule
		\emph{28-Pre}  & 60.0 & 380 & 30.0 & 30.0 & 15.0 & 0.48 & 170.65 &  12.00 &  6.75\\
		\emph{28-Post} & 55.0 & 380 & 30.0 & 30.0 & 15.0 & 0.48 & 170.65 & 12.00 & 6.75 \\
		\emph{44-Pre}  & 90.0 & 400 & 50.0 & 50.0 & 15.0 & 0.48 & 166.65 & 13.33 & 7.42\\
		\bottomrule
	\end{tabular}
	\caption{Fitted parameters for EM Model.}
	\label{tab:_EM_param_fit}
\end{table}

%% tab4
\begin{table}[htb]\centering
\ra{1.3}
		\begin{tabular}{@{}lccr@{}}\toprule
			\textbf{Name} & \textbf{Variable} & \textbf{Units} & \textbf{Expression} / \textbf{Value}\\
			\hline
			Velocity & $\vec u_\mathrm{f}$ & $\si{\metre\per\second}$  & $-$ \\
			Pressure & $p_\mathrm{f}$ & $\si{\pascal}$ & $-$ \\
			Fluid stress tensor & $\tens \sigma_\mathrm{f}(\vec u_\mathrm{f}, p_\mathrm{f})$ & $\si{\pascal}$ & $-p_\mathrm{f} \tens I + 2 \mu_\mathrm{f}\tens\varepsilon(\vec u_\mathrm{f})$ \\
			Fluid density  & $\rho_\mathrm{f}$ & $\si{\kilogram\per\cubic\metre}$ & $\num{1060}$ \\
			Dynamic viscosity& $\mu_\mathrm{f}$ & $\si{\pascal\second}$ & $\num{0.004}$ \\
			\bottomrule
		\end{tabular}
\caption{Physical parameters for the Navier--Stokes equations with their respective ranges and units. $\tens\varepsilon(\vec u_\mathrm{f})$ denotes the symmetric gradient defined as $\frac{1}{2}\left(\nabla_{\vec x} \vec u_\mathrm{f} + \nabla_{\vec x} \vec u_\mathrm{f}^\top\right)$}
\label{tab:parameters_navier_stokes}
\end{table}

%% tab5
\begin{table}[htb]

	\ra{1.5}
	\centering
	\begin{tabular}{@{}lcccccc@{}}
		\toprule
		& \multicolumn{6}{c}{\textbf{EM Comparison}}  \\
		\hline
		& $\mathrm{EDV}_{\mathrm{cl},\mathrm{sim}}$ & $\mathrm{ESV}_{\mathrm{cl},\mathrm{sim}}$ & $\mathrm{SV}_\mathrm{cl,sim}$ & $\mathrm{EF}_\mathrm{cl,sim}$& $\mathrm{CO}_\mathrm{cl,sim}$ & $\mathrm{P}_\mathrm{cl,sim}^\mathrm{sys}$\\
& $\si{\milli\litre}$ & $\si{\milli\litre}$ & $\si{\milli\litre}$ & $\si{\percent}$ & $\si{\milli\litre\per\second}$ &
          $\si{\mmHg}$ \\
%		& $\si{\kPa}$ & $\si{\ms}$ & $-$ & $-$ & $\si{\ms}$ & $\si{\kPa}$ & $\si{\kilo\Pa \milli\second\per\milli\litre}$  &  $\si[per-mode=symbol]{\kilo\Pa \milli\second\per\milli\litre}$  & $\si{\milli\litre\per\kilo\Pa}$ \\
		\midrule
		\emph{28-Pre}  & $\sfrac{88.16}{87.47}$ & $\sfrac{30.62}{31.02}$ & $\sfrac{57.54}{57.14}$ &
                $\sfrac{65.27}{64.81}$ & $\sfrac{87.46}{86.85}$ & $\sfrac{146.037}{139.362}$ \\
		\emph{44-Pre}  & $\sfrac{91.68}{91.67}$ & $\sfrac{31.59}{30.95}$ & $\sfrac{60.10}{60.72}$ &
                $\sfrac{65.54}{66.24}$ & $\sfrac{76.31}{76.32}$ & $\sfrac{158.413}{135.236}$\\
		\hline
		rel. error $[\si{\percent}]$  & $\sfrac{0.78}{0.01}$ & $\sfrac{1.3}{2.0}$ & $\sfrac{0.69}{1.03}$ & $\sfrac{0.70}{1.07}$ & $\sfrac{0.69}{0.013}$ & $\sfrac{4.57}{14.63}$ \\
		\bottomrule
	\end{tabular}
	\caption{Comparison of clinical indicators and indicators computed from simulation for the EM models.}
	\label{tab:_EM_validation}
\end{table}

%% tab6
\begin{table}[htb]
\ra{1.3}
\centering
\begin{tabular}{@{}lcccc|ccccc@{}}\toprule
& \multicolumn{4}{c|}{\textbf{Electromechanics Model}} & \multicolumn{5}{c}{\textbf{CFD Model} } \\
& NE & NV & $h$ [\si{\micro\meter}] & DOF & NE & NV & $h$ [\si{\micro\meter}] & DOFU & DOFP \\
\midrule
\emph{28-Pre} & 747266 & 167509 & 897 & 502527 & 1943060 & 352006 & 746.5 & 1056018 & 352006 \\
\emph{28-Post} & 632635 & 149174 & 954 & 447522 & 7405128 & 1294264 & 531.6 & 3882792 & 1294264 \\
\emph{44-Pre} & 727194 & 168804 & 997 & 506412 & 2285005 & 412728 & 717 & 1238184 & 412728 \\
\bottomrule
\end{tabular}
\caption{Discretization details for the studied cases. Shown are the number of elemens (NE), number of vertices (NV), average edge length $h$ in \si{\micro\meter}, degrees of freedom for displacement (DOF), degrees of freedom for velocity (DOFU), degrees of freedom for pressure (DOFP).}
\label{tab:meshes}
\end{table}

%% tab5
\begin{table}[htb]
\ra{1.3}
\centering
\begin{tabular}{@{}lcccc@{}}
\toprule
& \multicolumn{4}{c}{\textbf{Outlet}}  \\
& DCA & BCA & RSC & LSC  \\
\midrule
  $R~[\si{\kPa\ms\per\ml}]$ & $590.46$ & $264.6$ & $1058.24$ & $1058.24$\\
  $Z~[\si{\kPa\ms\per\ml}]$ & $29.52$ & $13.23$ & $52.91$ &$52.91$\\
  $C~[\si{\ml\per\kPa}]$ & $1.69$ & $3.78$ & $0.944$ &$0.944$\\
\bottomrule
\end{tabular}
\caption{Windkessel parameters for case \emph{28-Pre}.}
\label{tab:WK28}
\end{table}

%% tab6
\begin{table}
\ra{1.3}
\centering
\begin{tabular}{@{}lcccc@{}}
\toprule
& \multicolumn{4}{c}{\textbf{Outlet}}  \\
& DCA & BCA & RSC & LSC  \\
\midrule
  $R~[\si{\kPa\ms\per\ml}]$ & $2480.07$ & $276.01$ & $466.23$ & $7440.2$\\
  $Z~[\si{\kPa\ms\per\ml}]$ & $124.003$ & $13.9$ & $23.31$ & $372.01$\\
  $C~[\si{\ml\per\kPa}]$ & $0.403$ & $3.62$ & $2.14$ & $0.134$\\
\bottomrule
\end{tabular}
\caption{Windkessel parameters for case \emph{44-Pre}.}
\label{tab:WK44}
\end{table}

%% tab7
\begin{table}[htb]
\ra{1.3}
\centering
\begin{tabular}{@{}lccc|ccc@{}}
\toprule
\multicolumn{7}{c}{\textbf{Flux Comparison}} \\
\hline
& & \multicolumn{2}{c|}{\emph{28-Pre}} & \multicolumn{3}{c}{\emph{44-Pre}}  \\
& Unit & $\varepsilon_\mathrm{DCA}$ & $\varepsilon_\mathrm{ASC}$ & $\varepsilon_\mathrm{ASC}$ & $\varepsilon_\mathrm{BCA}$ & $\varepsilon_\mathrm{LSC}$  \\
\midrule
  $Q_\mathrm{peak,sim}$ & $\si{\milli\litre\per\second}$ & $85.5073$ & $286.056$ & $316.713$ & $160.493$ & $132.540$\\
  $Q_\mathrm{peak,cl}$ & $\si{\milli\litre\per\second}$ & $70.3071$ & $290.719$ & $352.114$ & $171.571$ & $109.290$\\
  rel. error & $\si{\percent}$ & $21.62$ & $1.604$ & $10.054$ & $6.46$ & $21.27$\\
\bottomrule
\end{tabular}
\caption{Comparison of clincal estimated flow rates and simulated flow rates through the various planes for cases \emph{28-Pre} and \emph{44-Pre}.}
\label{tab:flux_comp}
\end{table}

\clearpage
\section*{Figures}

%%fig1

%%fig2

%%fig3

%%fig4

%%fig5

%%fig6

%%fig7

%%fig8

%%fig9

\end{document}